\documentclass[12pt, a4paper, twoside]{article}

\usepackage{latexsym}
\usepackage{mathtools}
\usepackage{amsfonts,amssymb}
\usepackage{mathrsfs}
\usepackage{dsfont}
\usepackage[all]{xy}	

\usepackage{natbib}

\usepackage{bm}	

\usepackage{nicefrac} 
\usepackage{booktabs}	
\usepackage{setspace}	

\allowdisplaybreaks

\def\bSig\mathbf{\Sigma}

\newcommand{\beqo}{\[}
\newcommand{\eeqo}{\]}
\newcommand{\beqq}{\begin{equation}}
\newcommand{\eeqq}{\end{equation}}

\let\oldsqrt\sqrt
\def\sqrt{\mathpalette\DHLhksqrt}
\def\DHLhksqrt#1#2{%
\setbox0=\hbox{$#1\oldsqrt{#2\,}$}\dimen0=\ht0
\advance\dimen0-0.2\ht0
\setbox2=\hbox{\vrule height\ht0 depth -\dimen0}%
{\box0\lower0.4pt\box2}}

\newcommand{\R}{\mathbb{R}}

\newcommand{\calI}{\mathcal{I}}
\newcommand{\calJ}{\mathcal{J}}

\newcommand{\calB}{\mathcal{B}}
\newcommand{\calC}{\mathcal{C}}
\newcommand{\calD}{\mathcal{D}}

\newcommand{\calL}{\mathcal{L}}
\newcommand{\calN}{\mathcal{N}}

\newcommand{\simiid}{\stackrel{\text{i.i.d.}}{\sim}}

\newcommand{\veps}{{\varepsilon}}

\newcommand{\te}{{\theta}}

\newcommand{\bveps}{\bm \varepsilon}

\newcommand{\btheta}{\bm \theta}
\newcommand{\bmu}{\bm \mu}
\newcommand{\bnu}{\bm \nu}
\newcommand{\balpha}{\bm \alpha}
\newcommand{\bbeta}{\bm \beta}
\newcommand{\bgamma}{\bm \gamma}
\newcommand{\bomega}{\bm \omega}
\newcommand{\bxi}{\bm \xi}
\newcommand{\beeta}{\bm \eta}

\newcommand{\bsigma}{\bm \sigma}

\newcommand{\hattheta}{\widehat{\bm \theta}}

\newcommand{\hatbeta}{\widehat{\bm \beta}}

\newcommand{\hateta}{\widehat{\bm \eta}}

\newcommand{\bnull}{{\bm 0}}

\newcommand{\bU}{{\bm{U}}}
\newcommand{\bu}{{\bm{u}}}

\newcommand{\bx}{{\bm{x}}}
\newcommand{\by}{{\bm{y}}}
\newcommand{\bY}{{\bm{Y}}}

\newcommand{\E}{\mbox{E}}

\newcommand{\Var}{\mbox{Var}}

\newcommand{\Fzi}{g_z^{-1}}
\newcommand{\itau}{g_{\te}}

\newcommand{\vN}[2]{\mathcal{N}\left\{#1,#2\right\}}
\newcommand{\mvN}[3]{\mathcal{N}_{#1}\left\{#2,#3\right\}}

\newcommand{\vst}[4]{\text{skew-}t\left\{#1,#2,#3,#4\right\}}
\newcommand{\vU}[2]{\mathcal{U}\left(#1,#2\right)}

\newcommand{\cdfst}[5]{F_{\text{skew-}t}\left\{#5 \sep #1,#2,#3,#4\right\}}
\newcommand{\qst}[5]{F_{\text{skew-}t}^{-1}\left(#5 \sep #1,#2,#3,#4\right)}

\newcommand{\ov}{\overline}
\newcommand{\wh}{\widehat}
\newcommand{\wt}{\widetilde}

\newcommand{\sep}{\,|\,}
\newcommand{\sepi}{;}

\newcommand{\half}{\frac{1}{2}}

\newcommand{\iN}[2]{#1=1,\ldots,#2}
\newcommand{\xn}[2]{#1_1,\ldots,#1_{#2}}

\newcommand{\sumtN}{\sum_{t=1}^N}

\newcommand{\T}{^\top}	

\newcommand{\lnf}[1]{\ln\left(#1\right)}

\newcommand{\vrtx}{\mathcal{V}}	
\newcommand{\edge}{\mathcal{E}}	
\newcommand{\tree}{\mathcal{T}}	

\newcommand{\vine}{\mathscr{V}}	
\newcommand{\Cop}[2]{C_{#1}\left\{#2\right\}}	
\newcommand{\cop}[2]{c_{#1}\left\{#2\right\}}	
\newcommand{\Lik}[3]{\calL_{#1}\left(#2\sep#3\right)}	
\renewcommand{\ll}[3]{\ell_{#1}\left(#2\sep#3\right)}	

\usepackage[figuresright]{rotating}

\usepackage[labelsep=period]{caption}

\begin{document}

\title{\textbf{R-vine Models for Spatial Time Series with an Application to Daily Mean Temperature}}

\author{by\\
\\
Tobias Michael Erhardt$^*$, Claudia Czado, and Ulf Schepsmeier\\
Zentrum Mathematik\\
Technische Universit\"at M\"unchen\\
Boltzmannstr. 3, 85748 Garching, Germany\\
\\
$^*$tobias.erhardt@tum.de}

\maketitle


\begin{abstract}

We introduce an extension of R-vine copula models for the purpose of spatial dependency modeling and model based prediction at unobserved locations. The newly derived spatial R-vine model combines the flexibility of vine copulas with the classical geostatistical idea of modeling spatial dependencies by means of the distances between the variable locations. In particular the model is able to capture non-Gaussian spatial dependencies. For the purpose of model development and as an illustration we consider daily mean temperature data observed at $54$ monitoring stations in Germany. We identify a relationship between the vine copula parameters and the station distances and exploit it in order to reduce the huge number of parameters needed to parametrize a $54$-dimensional R-vine model needed to fit the data. The new distance based model parametrization results in a distinct reduction in the number of parameters and makes parameter estimation and prediction at unobserved locations feasible. The prediction capabilities are validated using adequate scoring techniques, showing a better performance of the spatial R-vine copula model compared to a Gaussian spatial model.\\


\end{abstract}

%

\clearpage

\section[Introduction]{Introduction}\label{sec:intro}

Comprehension of the earth's climate system is of vital interest to every aspect of human life. Recently the class of vine copulas has captured attention as a flexible class to model high dimensional dependencies \citep*[see][and reference therein]{czado10, czado13, kurowickacooke06, kurowickajoe11}. We present a new vine copula based approach for the spatial modeling of climatic time series. Utilization of available spatial information will lead to a distinct reduction in the number of parameters needed to parametrize the high dimensional (spatial) regular vine (R-vine) copula model. Model selection, estimation and a prediction method at arbitrary locations will be developed.

Different approaches to model spatial dependencies can be found in the literature. To name a few, we start with \citet*[][]{stahl06}, who compare different deterministic and stochastic spatial interpolation models with an application to daily minimum and maximum temperatures.
\citet*[][]{benth07} introduce a Gaussian random field based spatial-temporal model for daily temperature averages and
\citet*[][]{hu13} use systems of stochastic partial differential equations to model the dependence of temperature and humidity spatially. These two approaches are restricted to model Gaussian dependencies, whereas our methodology allows for non-Gaussian dependencies.
Another vine copula based modeling approach in a spatio-temporal framework is presented by \citet*[][]{graeler12}. Their approach combines geostatistical methods with copulas to model the spatial dependencies in the first vine copula trees. In contrast our new approach also allows to include spatial covariates other than distance, to model the spatial dependencies in all vine copula trees.

The customary tool applied for dependency modeling are multivariate Gaussian distributions. However these distributions are not appropriate to model any data, since they require symmetry and do not allow for extreme dependency. Therefore we apply vine copula models, which are designed to overcome these limitations. Copulas are $d$-dimensional distribution functions on $[0,1]^d$ with uniform margins. They can be understood as a tie between a multivariate distribution function $F$ and its marginals ($\xn{F}{d}$) and capture all dependency information \citep[see][]{sklar59}. In particular it holds $F(\by)=C\left(F_1(y^1),\ldots,F_d(y^d)\right)$, where $\by=(y^1,\ldots,y^d)\T$ is the realization of a random vector $\bY\in\R^d$. Vine copulas, are constructions of $d$-dimensional copulas built on bivariate copulas only. They are well understood and easy to compute \citep[see][]{aas09, cdvine13, dissmann13}. A short introduction to R-vines will be given in Section \ref{sec:rvines}.

We develop our approach for daily mean temperature time series collected over the period 01/01/2010-12/31/2012 by the German Meteorological Service (Deutscher Wetterdienst) (Section \ref{sec:margins}). The common modeling of all marginal distributions is discussed in Section \ref{sec:margins}. It captures seasonality effects and temporal dependencies of the time series. Spatially varying parameters allow to approximate these effects and dependencies at unobserved locations.

The main contribution is the development of a new vine copula based spatial dependency model introduced in Section \ref{sec:svines}. It relies on a reparametrization of an R-vine copula model, which exploits the relationship between the model parameters and the available spatial information. Different model specifications based on distances and elevation differences were considered in \citet[][]{erhardt13}, the most promising one is highlighted here. Maximum likelihood estimation is followed by model based prediction at a new location.

A geostatistical model is developed in Section \ref{sec:classical} and used for comparison. The resulting model evaluation is conducted in Section \ref{sec:appl}. A validation data set for $19$ additional locations allows to calculate adequate scores, based on which the quality of the predictions can be compared. The outcome of our investigations is discussed in Section \ref{sec:discuss}.

An application of the presented methodology, not only in the area of climatic research, but also in other areas which require the modeling of spatial dependencies, is possible. This would require the development of appropriate marginal models, tailored to the characteristics of the respective data. Application of the methods to the modeling of monitoring systems such as pollutants and biomass can be envisioned.


\section[Regular vine copula models]{Regular vine copula models}\label{sec:rvines}

\emph{Vine copulas} in general were introduced by \citet[][]{bedfordcooke01,bedfordcooke02} and trace back to ideas of \citet[][]{joe96}.
They are build using a cascade of $\nicefrac{d(d-1)}{2}$ bivariate copulas, called \emph{pair copulas}. This cascade is identified using a set of nested trees called a regular vine tree sequence or short \emph{regular vine} (R-vine).
In particular the R-vine tree sequence $\vine=(\xn{\tree}{d-1})$ satisfies the following conditions \citep[see][]{bedfordcooke01}:
\begin{enumerate}
	\item $\tree_1=(\vrtx_1,\edge_1)$ is a tree with vertices $\vrtx_1=\{1,\ldots,d\}$ and edge set $\edge_1$.
	\item $\tree_l=(\vrtx_l,\edge_l)$ is a tree with vertices $\vrtx_l=\edge_{l-1}$ and edge set $\edge_l$, for all $l=2,\ldots,d-1$.
	\item For all vertex pairs in $\vrtx_l$ connected by an edge $e\in\edge_l$, $l=2,\ldots,d-1$, the corresponding edges in $\edge_{l-1}$ have to share a common vertex (\emph{proximity condition}).
\end{enumerate}
\citet[][]{aas09} were the first to develop statistical inference for non-Gaussian pair copulas.

Our notation of the vine edges will follow \citet{czado10}. An edge $e \in \edge_l$, $\iN{l}{d-1}$, will be denoted by $i(e),j(e);\calD_e$, where $i(e)<j(e)$ make up the \emph{conditioned set} $\calC_e=\{i(e), j(e)\}$ and $\calD_e$ is called \emph{conditioning set}. An example in five dimensions is given in Figure \ref{fig:rvine}. It depicts the four nested trees of an R-vine tree sequence $\vine=(\xn{\tree}{4})$.

Next we introduce the link of an R-vine tree sequence $\vine$ to the multivariate copula distribution of some random vector $\bU=(U^1,\ldots,U^d) \in [0,1]^d$ with $U^1,\ldots,U^d \sim \vU01$. We define the set $\calB\coloneqq\left\{C_{i(e),j(e);\calD_e}: e\in\edge_l, \iN{l}{d-1}\right\}$ of bivariate copulas $C_{i(e),j(e);\calD_e}$ corresponding to the R-vine edges $e\in\edge_l$, $\iN{l}{d-1}$. These copulas may be parametrized by several parameters, depending on their pair copula family $b_{i(e),j(e);\calD_e}$. For an overview of frequently used bivariate copula families we refer to \citet{cdvine13}. Moreover we define $\bu^{\calI}	\coloneqq \left\{u^k: k\in\calI\right\}$ for arbitrary index sets $\calI\subseteq\{1,\ldots,d\}$. This allows to formulate the vine copula density of $\bU$ corresponding to the R-vine tree sequence $\vine$
as
\beqq\label{eq:vcdensity}
		c_{1,\ldots,d}(\bu)=\prod_{l=1}^{d-1} \prod_{e\in\edge_l} \cop{i(e),j(e);\calD_e}{C_{i(e)|\calD_e}(u^{i(e)}\sep\bu^{\calD_e}),C_{j(e)|\calD_e}(u^{j(e)}\sep\bu^{\calD_e})},
\eeqq
	where $\cop{i(e),j(e);\calD_e}{\cdot,\cdot}$ are the densities corresponding to the bivariate copulas $C_{i(e),j(e);\calD_e}\in\calB$. See \citet[][]{bedfordcooke01} for derivation of \eqref{eq:vcdensity}. To evaluate such a density, we need to calculate the so called \emph{transformed variables} $C_{i(e)|\calD_e}(u^{i(e)}\sep\bu^{\calD_e})$ and $C_{j(e)|\calD_e}(u^{j(e)}\sep\bu^{\calD_e})$. Here $C_{i(e)|\calD_e}$ and $C_{j(e)|\calD_e}$ are conditional distributions obtained from $C_{i(e),j(e);\calD_e}$. The calculation is performed recursively according to \citet{joe96}, using the formula
\beqq\label{eq:joe}
	C_{k|\calJ}(u^k \sep \bu^{\calJ}) =	\frac{\partial\ \Cop{kl;\calJ_{-l}}{C_{k|\calJ_{-l}}(u^k \sep \bu^{\calJ_{-l}}), C_{l|\calJ_{-l}}(u^l \sep \bu^{\calJ_{-l}})}}{\partial C_{l|\calJ_{-l}}(u^l \sep \bu^{\calJ_{-l}})},
\eeqq
where $k,l\in\{1,\ldots,d\}$, $k\neq l$, $\{l\}\subset\calJ\subset\{1,\ldots,d\}\backslash\{k\}$ and $\calJ_{-l}\coloneqq\calJ\backslash\{l\}$. We implicitly made a \emph{simplifying assumption}, that the copula distributions in $\calB$ do not depend on the conditioning value $\bu^{\calD_e}$ other than through its arguments given in \eqref{eq:vcdensity}. 

In our spatio-temporal setting the data $y_t^s$, $\iN{s}{d}$, $\iN{t}{N}$, is not restricted to the unit hypercube $[0,1]^d$ and does not necessarily have uniformly distributed margins. For that reason the data has to be transformed to so called \emph{copula data} $u_t^s\sim\vU01$, $\iN{s}{d}$, $\iN{t}{N}$, before vine copula models can be applied. We consider a regression model $Y_t^s=g(t, \bx^s; \bbeta) + \veps_t^s$, $\veps_t^s \sim F^s$, with spatial covariates $\bx^s$, to adjust for spatial as well as seasonality effects and temporal dependencies. The resulting residuals $\wh\veps_t^s \coloneqq y_t^s - g(t, \bx^s; \wh\bbeta)$, $\iN{t}{N}$, are now approximately independent for each location $\iN{s}{d}$. We transform these residuals by their respective parametric marginal distribution functions $F^s$, i.e. we calculate $u_t^s \coloneqq F^s(\wh\veps_t^s)$. This transformation is called the probability integral transform. We prefer to use parametric probability integral transformations \citep[see][]{joe96b} over empirical rank transformations \citep*[proposed for example by][]{genest95}, since we are interested in predictions on the original scale using the proposed marginal models.

In Section \ref{sec:svines} we will use \emph{truncated} R-vines \citep*{brechmann12}. Truncation after a certain level $k<d-1$ means that $C_{i(e),j(e);\calD_e}$ are chosen to be independence copulas for all edges $e\in\edge_l$, $k<l<d$.

\section[A Marginal Model for Daily Mean Temperatures]{A Marginal Model for Daily Mean Temperatures}\label{sec:margins}

The data set consists of daily mean temperature data in $^{\circ}\mathrm{C}$ collected over the period 01/01/2010-12/31/2012 by the German Meteorological Service (Deutscher Wetterdienst) at $73$ selected observation stations across Germany. The data set is split into a training ($\iN{s}{54}$) and a validation data set ($s=55,\ldots,73$, see Table \ref{tab:scores} for stations selected).
Hence we build our models on $d=54$ times $N=1096$ observations $y_t^s$ of daily mean temperatures, which are considered
as realizations of random variables $Y_t^s$ ($\iN{t}{N}$, $\iN{s}{d}$).

\begin{figure}[htb]
	\centering
		\includegraphics[width=0.48\textwidth]{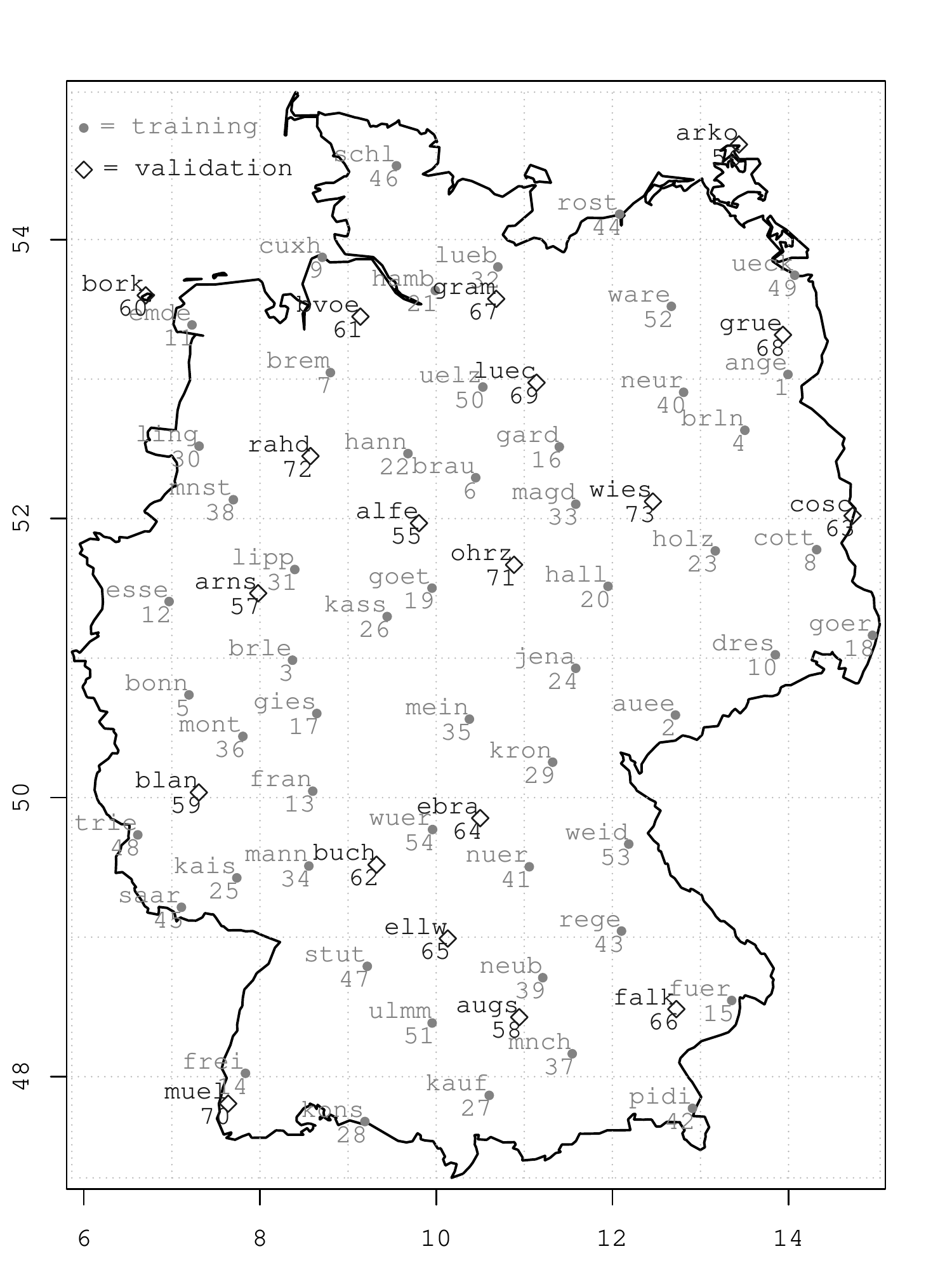}
	\caption[The $73$ observation stations across Germany.]{The $73$ observation stations across Germany with ID and respective short name: Training data ($\iN{s}{54}$) and validation data ($s=55,\ldots,73$).}
	\label{fig:stations}
\end{figure}

Lists with detailed information about the location (longitude ($x_{\text{lo},s}$), latitude ($x_{\text{la},s}$) and elevation ($x_{\text{el},s}$)) and the names of all $73$ observation stations are given in Table 3.1 and 5.9 of \citet{erhardt13}. The location of the stations in Germany is illustrated in Figure \ref{fig:stations}.

For vine copula based models, we need to transform our data to copula data. For this, we use the marginal model of \citet[][Chapter 3]{erhardt13}, which is a tailor-made model for the marginal mean temperatures at arbitrary locations in Germany. To ensure homoscedasticity, i.e. $\Var(\veps_t^s)=\sigma^2>0$, $\iN{t}{N}$, $\iN{s}{d}$, the model considers appropriately weighted observations $\wt Y_t^s \coloneqq \nicefrac{Y^s_t}{\sqrt{\wh w_t}}$. Raw weights $\wt w_t$, $\iN{t}{N}$, obtained as the sample variances
$\wt w_t \coloneqq \frac{1}{d-1}\sum_{s=1}^d \left(y_t^s - \ov y_t\right)^2$, where $\ov y_t \coloneqq \frac{1}{d}\sum_{s=1}^d y_t^s$, $\iN{t}{N}$,
are smoothed by means of least squares. This results in the \emph{smoothed weights}
$\wh w_t \coloneqq \exp\left\{q(t;\wh\balpha)\right\}$.
Here $q$ is chosen to be a polynomial in $t$ of degree nine.

\subsection{Model Components}\label{sec:mod:comp}
We now outline the most important features of the different model components. For details we refer to \citet[][Chapter 3]{erhardt13}.

\paragraph{Annual seasonality}\label{sec:season}

Yearly temperature fluctuations can be captured by sine curves of the form $\lambda\sin(\omega t + \delta)$,
parametrized by $\lambda$ (amplitude), $\omega$ (angular frequency) and $\delta$ (phase shift). A substitution of these parameters, inspired by \citet[][]{simmons90}, leads to the linear model component
$\beta_{\text{s}}\sin(\omega t) + \beta_{\text{c}}\cos(\omega t)$,
where $\omega$ is set to $\nicefrac{2\pi}{365.25}$, due to the annual context.

\paragraph{Autoregression}\label{sec:ar}

Temporal dependence is eliminated by the inclusion of an autoregression component of the form
$\sum_{j=1}^{q} \gamma_j Y_{t-j}$
into the marginal model. Investigations show that the choice of $q=3$ lagged responses as additional covariates is appropriate.

\paragraph{Skew-$t$ distributed errors}\label{sec:skew:dist}

Detailed investigations showed that skew-$t$ distributed errors
$\xn{\veps}{N}\simiid\vst{\xi}{\omega}{\alpha}{\nu}$,
are appropriate, since they are able to capture the observed skewness and heavy tails. The parametrization of \citet[][]{azzalini03} is utilized, which results in the probability density function
\beqq\label{eq:skewt}
	f_{\text{skew-}t}\left(x; \xi, \omega, \alpha, \nu\right) = \frac{2}{\omega} t_{\nu}(\wt x) T_{\nu+1}\left\{\alpha\wt x\left(\frac{\nu+1}{\nu+\wt x^2}\right)^{\nicefrac{1}{2}}\right\},
\eeqq
where $\wt x\coloneqq\nicefrac{(x-\xi)}{\omega}$. Here $t_{\nu}$ is the density and $T_{\nu+1}$ the cumulative distribution function of a usual, univariate Student-$t$ distribution with $\nu$ and $\nu+1$ degrees of freedom, respectively. Whereas the parameters $\xi$, $\omega$ and $\alpha$ can be interpreted as location, scale and shape parameter, respectively, $\nu$ denotes the degree of freedom parameter of the skew-$t$ distribution.

\paragraph{Aggregated parameters}\label{sec:agg:par}

The parameters of the previously described model components are replaced by polynomial structures, in order to account for spatial variation in the temperatures depending on longitude, latitude and elevation. We call them aggregated or spatially varying parameters.

\subsection[The Marginal Model]{The Marginal Model}\label{sec:tmm}

The marginal model for the daily mean temperatures is given as
\beqq\label{eq:jointmargmod}
	\wt Y_t^s = \mu_t^s + \veps_t^s, \quad	\veps_t^s \sim \vst{\xi(s)}{\omega(s)}{\alpha(s)}{\nu(s)}, \quad \iN{t}{N}, \,\iN{s}{d},
\eeqq
with mean function
\begin{align*}
		\mu_t^s	& \coloneqq g\left(t, \wt Y_{t-1}^s, \wt Y_{t-2}^s, \wt Y_{t-3}^s, x_{\text{el},s}, x_{\text{lo},s}, x_{\text{la},s} \sepi \bbeta\right) \\*
						& \coloneqq \beta_0(s) + \beta_{\text{s}}(s) \sin\left(\frac{2\pi t}{365.25}\right)
								+ \beta_{\text{c}}(s) \cos\left(\frac{2\pi t}{365.25}\right)	+ \gamma_1(s) \wt Y_{t-1}^s + \gamma_2(s) \wt Y_{t-2}^s + \gamma_3(s) \wt Y_{t-3}^s,
\end{align*}
where the spatially varying parameters are divided into the \emph{aggregated intercept and seasonality parameters}
\begin{align*}
	\beta_{0}(s)	& \coloneqq \beta_{00} + \beta_{011} x_{\text{el},s} + \beta_{031} x_{\text{la},s},	\\
	\beta_{\text{s}}(s)	& \coloneqq \beta_{\text{s}0} + \sum_{j=1}^4{\beta_{\text{s}1j} x_{\text{el},s}^j} + \beta_{\text{s}21} x_{\text{lo},s} + \sum_{l=1}^6{\beta_{\text{s}3l} x_{\text{la},s}^l},	\\
	\beta_{\text{c}}(s)	& \coloneqq \beta_{\text{c}0} + \sum_{j=1}^6{\beta_{\text{c}1j} x_{\text{el},s}^j} + \sum_{k=1}^2{\beta_{\text{c}2k} x_{\text{lo},s}^k} + \beta_{\text{c}31} x_{\text{la},s},
\end{align*}
the \emph{aggregated autoregression parameters}
\begin{align*}
	\gamma_{1}(s)	& \coloneqq \gamma_{10} + \gamma_{111} x_{\text{el},s} + \sum_{k=1}^2{\gamma_{12k} x_{\text{lo},s}^k} + \sum_{l=1}^6{\gamma_{13l} x_{\text{la},s}^l},	\\
	\gamma_{2}(s)	& \coloneqq \gamma_{20} + \gamma_{211} x_{\text{el},s} + \sum_{k=1}^2{\gamma_{22k} x_{\text{lo},s}^k} + \sum_{l=1}^6{\gamma_{23l} x_{\text{la},s}^l},	\\
	\gamma_{3}(s)	& \coloneqq \gamma_{30} + \sum_{k=1}^4{\gamma_{32k} x_{\text{lo},s}^k} + \sum_{l=1}^7{\gamma_{33l} x_{\text{la},s}^l},
\end{align*}
and the \emph{aggregated skew-$t$ parameters}
\begin{align*}
	\xi(s)	& \coloneqq \xi_{0} + \xi_{11} x_{\text{el},s} + \sum_{k=1}^2{\xi_{2k} x_{\text{lo},s}^k} + \xi_{31} x_{\text{la},s},	\\
	\omega(s)	& \coloneqq \exp\left\{\omega_{0} + \sum_{j=1}^3{\omega_{1j} x_{\text{el},s}^j} + \omega_{21} x_{\text{lo},s} + \sum_{l=1}^6{\omega_{3l} x_{\text{la},s}^l}\right\},	\\
	\alpha(s)	& \coloneqq \alpha_{0} + \sum_{j=1}^4{\alpha_{1j} x_{\text{el},s}^j} + \sum_{k=1}^2{\alpha_{2k} x_{\text{lo},s}^k} + \alpha_{31} x_{\text{la},s},	\\
	\nu(s)	& \coloneqq \exp\left\{\nu_{0} + \sum_{j=1}^2{\nu_{1j} x_{\text{el},s}^j} + \sum_{k=1}^2{\nu_{2k} x_{\text{lo},s}^k} + \sum_{l=1}^4{\nu_{3l} x_{\text{la},s}^l}\right\}.
\end{align*}
The model parameters are summarized in the two vectors
$\bbeta \coloneqq \left(\bbeta_0\T, \bbeta_{\text{s}}\T, \bbeta_{\text{c}}\T, \bgamma_1\T, \bgamma_2\T, \bgamma_3\T\right)\T\in\R^{57}$ and $\beeta \coloneqq \left(\bxi\T, \bomega\T, \balpha\T, \bnu\T\right)\T\in\R^{33}$.

\subsection[Marginal Model Parameter Estimation]{Marginal Model Parameter Estimation}\label{sec:mmpe}

The parameter estimation follows a two step approach. In a first step the
parameters $\bbeta$ are estimated by least-squares estimation. Thereafter we calculate the raw residuals $\wh\veps_t^s \coloneqq \wt y_t^s - \wh\mu_t^s$, $t=4,\ldots,N$, $\iN{s}{d}$. Note that they cannot be computed for $t=1,2,3$, due to the autoregression of $\wt y_t^s$ onto the three previous points in time. The residuals $\wh\veps_t^s$ are used to fit the skew-$t$ parameters $\beeta$ by maximization of the pseudo-likelihood
\beqo
	\Lik{\text{skew-}t}{\beeta}{\wh\bveps^1,\ldots,\wh\bveps^d} = \prod_{s=1}^{d} \prod_{t=4}^{N} f_{\text{skew-}t}\left\{\wh\veps_t^s; \xi(s), \omega(s), \alpha(s), \nu(s)\right\}.
\eeqo
This results in the vector
$\hattheta \coloneqq	\left(\hatbeta\T, \hateta\T\right)\T$
of marginal parameter estimates for Model \eqref{eq:jointmargmod}.

\subsection{Transformation to Copula Data}\label{sec:copdat}

Finally we use the fitted Model \eqref{eq:jointmargmod} to transform our data to copula data, i.e. we transform our original time series $y_1^s,\ldots,y_N^s$, to $u_4^s,\ldots,u_N^s \simiid \vU01$ for all $\iN{s}{d}$.
Since for all $\iN{s}{d}$ we modeled the errors $\veps_1^s,\ldots,\veps_N^s$ as i.i.d. skew-$t$ distributed with spatially varying parameters $\xi(s)$, $\omega(s)$, $\alpha(s)$ and $\nu(s)$, the desired copula data is obtained as
\beqo
	u_t^s \coloneqq	\cdfst{\wh\xi(s)}{\wh\omega(s)}{\wh\alpha(s)}{\wh\nu(s)}{\wh\veps_t^s},	\quad	t=4,\ldots,N, \,\iN{s}{d},
\eeqo
where $\cdfst{\xi}{\omega}{\alpha}{\nu}{\cdot}$ is the cumulative distribution function corresponding to \eqref{eq:skewt}.


\section[A Spatial R-vine Model for Daily Mean Temperatures]{A Spatial R-vine Model for Daily Mean Temperatures}\label{sec:svines}

For spatial data, the spatial arrangement of the data plays an important role with regard to dependency modeling. As climatic data such as temperature is measured at a large number of spatial locations,
we face a high dimensional problem. With rising dimensionality ordinary R-vine copula models become computationally infeasible since the number of parameters increases quadratically. Exploitation of spatial information in our new approach of a \emph{spatial R-vine copula model (SV)} allows to reduce the number of parameters significantly.

\subsection{Preliminary Analyses}\label{sec:prem:ana}

To develop a spatial R-vine model we consider the copula data $\bu^1,\ldots,\bu^d$ where $\bu^s=(u_{1}^s,\ldots,u_{N}^s)\T$ and $u_{1}^s,\ldots,u_{N}^s \simiid \vU01$ for all $\iN{s}{d}$, i.e. we have copula data series of length $N$ for $d$ different observation stations. From the quantities elevation, longitude and latitude we are able to calculate a
\emph{distance} $d_{i,j}$
and a
\emph{elevation difference} $e_{i,j}$
for each pair of observation stations $(i,j)$ with $1 \leq i < j \leq d$.

We will allow for one- and two-parametric pair-copula families, whose first and second copula parameters are denoted as $\te_{i,j;\calD_e}$ and $\nu_{i,j;\calD_e}$.
The corresponding Kendall's $\tau$'s are denoted as $\tau_{i,j}$ respectively $\tau_{i,j;\calD_e}$,
depending on whether they are calculated directly from the data or based on transformed variables in the trees $\tree_2, \tree_3, \ldots, \tree_{d-1}$ of the R-vine.

\begin{figure}[htb]
	\centering
		\includegraphics[width=1.00\textwidth]{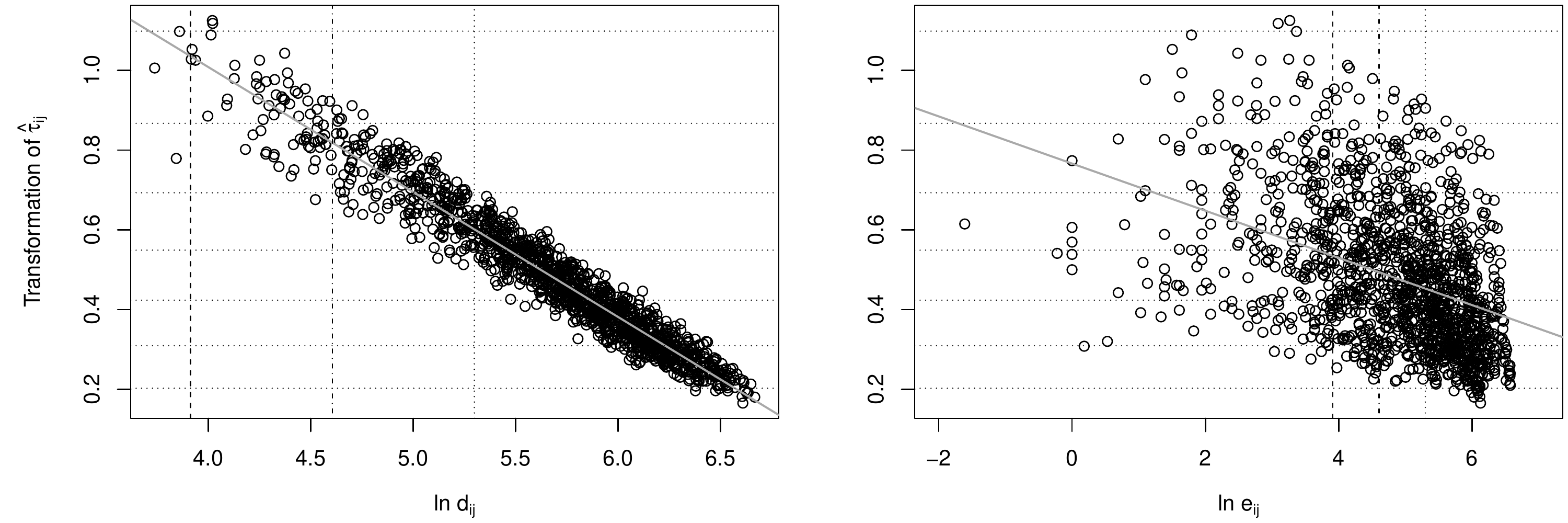}
	\caption{Relationship of Fisher z-transformed estimated Kendall's $\tau$'s $g_z(\wh\tau_{i,j})$ with log-distance $\ln\left(d_{i,j}\right)$ and log-elevation $\ln\left(e_{i,j}\right)$, respectively.}
	\label{fig:modtaureg}
\end{figure}

Returning to the mean temperature data set ($d=54$) we further investigate the spatial dependencies of the given variables. To this end we are interested in identifying a relationship between the dependence strength and the distance respective the elevation difference of observation station pairs. These relationships are illustrated in Figure \ref{fig:modtaureg}. For all $\nicefrac{d(d-1)}{2}=1431$ possible station pairs $(i,j)$, $1 \leq i < j \leq 54$, the empirical Kendall's $\tau$ values $\wh\tau_{i,j}$ are estimated, to quantify the dependence of these pairs. As they are restricted to $(-1,1)$, and since we want to take a linear model into consideration, we apply the Fisher z-transform
\beqq\label{eq:fisher}
	g_z(r) = \half \ln\left(\frac{1+r}{1-r}\right),\, r\in(-1,1),
\eeqq
first introduced by \citet[][]{fisher15}, to transform from $(-1,1)$ to $(-\infty,\infty)$.

The left panel of Figure \ref{fig:modtaureg} illustrates the Fisher z-transformed estimated Kendall's $\tau$'s against the logarithmized distances $\ln\left(d_{i,j}\right)$. Here, a distinct linear relationship can be observed. The right panel gives the respective plot against the logarithmized elevation differences $\ln\left(e_{i,j}\right)$. The observed linear relationship is not that distinct as for the distances.

The straight gray lines in both plots depict the regression line corresponding to the particular linear relationship. The horizontal lines help to identify the level of Kendall's $\tau$, whereas the vertical lines indicate the three distances of $50$, $100$ and $200$ kilometers and the three elevation differences of $50$, $100$ and $200$ meters, respectively.

A tree-wise analysis of an R-vine model fitted to the mean temperature data might lead to a deeper insight into the relationship of the R-vine copula parameters and the available spatial information. For these investigations we consider an R-vine truncated after tree ten, which allows for bivariate Gaussian ($\Phi$), Student-$t$ ($t$), Clayton (C), Gumbel (G) and Frank (F) copulas as pair copulas. Rotated versions of the Clayton and Gumbel copula are allowed in addition, to capture possible negative and asymmetric dependencies. The copula families are selected separately for each bivariate building block according to the Akaike information criterion. For more details on these copula families, copula rotation and copula selection we refer to \citet{cdvine13}. The R-vine tree structure is selected by tree-wise selection of maximum spanning trees, where Kendall's $\tau$'s are used as edge weights \citep[see][]{dissmann13}.
Application of a bivariate asymptotic independence test \citep[][]{genest07} in the copula family selection procedure yields a share of independence copulas of more than $50\%$ in all trees $\tree_l$ with $l\geq10$. Thus the truncation after level $10$  resulting in a significant reduction in the number of model parameters is reasonable.

Subsequently we modify our notation to indicate dependence on the edge $e\in\edge_l$. In particular cases we add a superscript $l\leq10$ to emphasize the respective tree number.

Table \ref{tab:SV:struct} summarizes the structure of the R-vine which we are going to investigate in more detail. We observe that the copula family which occurs most for the trees one to nine is the bivariate Student-$t$ copula family. It is the only two-parametric copula family under consideration. Further the number of other copula families increases with the tree number. In tree ten the Gumbel family is the dominating one. The Kendall's $\tau$'s in tree number two and higher are calculated based on transformed variables.
Further we observe from Table \ref{tab:SV:struct}, that the strong dependencies are already captured in tree one and that the association in higher trees scatters mostly between $-0.2$ and $0.3$, i.e. negative dependencies occur as well.

\begin{table}[htb]
\centering
\caption[Summary of the estimated structure of the truncated R-vine under consideration.]{Summary of the estimated structure of the truncated R-vine under consideration. Besides the numbers of the different copula families ($\Phi$=Gaussian, $t$=Student-$t$, C=Clayton, G=Gumbel, F=Frank pair copula) selected for each tree, the minimum and the maximum estimated Kendall's $\tau$'s and the averages over the occurring estimated second copula parameters ($\ov{\wh\nu^l} \coloneqq \frac{1}{\# t}\sum_{e \in\edge_l^t}\wh\nu_{i(e),j(e);\calD_e}^l$, $\edge_l^t \coloneqq \{e \in\edge_l: b_{i(e),j(e);\calD_e}\,\text{is a Student-$t$ copula}\}$) are provided.}
\label{tab:SV:struct}
\begin{tabular}{lrrrrrrrr}
	\hline
	\hline
	tree ($l$) & \# $\Phi$ & \# $t$ & \# C & \# G & \# F & $\displaystyle{\min_{e\in\edge_l}(\wh\tau_{i(e),j(e);\calD_e}^l)}$ & $\displaystyle{\max_{e\in\edge_l}(\wh\tau_{i(e),j(e);\calD_e}^l)}$ & $\ov{\wh\nu^l}$ \\ 
  \hline
	1 & 0 & 53 & 0 & 0 & 0 & 0.591 & 0.809 & 7.659 \\ 
  2 & 1 & 38 & 1 & 7 & 5 & -0.153 & 0.317 & 9.908 \\ 
  3 & 1 & 35 & 4 & 6 & 5 & -0.222 & 0.356 & 10.837 \\ 
  4 & 2 & 23 & 5 & 13 & 7 & -0.180 & 0.300 & 11.873 \\ 
  5 & 1 & 21 & 9 & 9 & 9 & -0.154 & 0.278 & 11.843 \\ 
  6 & 5 & 20 & 5 & 12 & 6 & -0.114 & 0.285 & 12.967 \\ 
  7 & 6 & 15 & 10 & 10 & 6 & -0.193 & 0.239 & 13.378 \\ 
  8 & 5 & 18 & 4 & 7 & 12 & -0.098 & 0.193 & 14.888 \\ 
  9 & 7 & 15 & 6 & 8 & 9 & -0.066 & 0.279 & 14.373 \\ 
  10 & 3 & 10 & 8 & 16 & 7 & -0.128 & 0.246 & 14.105 \\ 
	\hline
	\textbf{Sum} & \textbf{31} & \textbf{248} & \textbf{52} & \textbf{88} & \textbf{66} &  &  &  \\ 
	\hline
\end{tabular}
\end{table}

\begin{figure}[htb]
	\centering
		\includegraphics[width=0.48\textwidth]{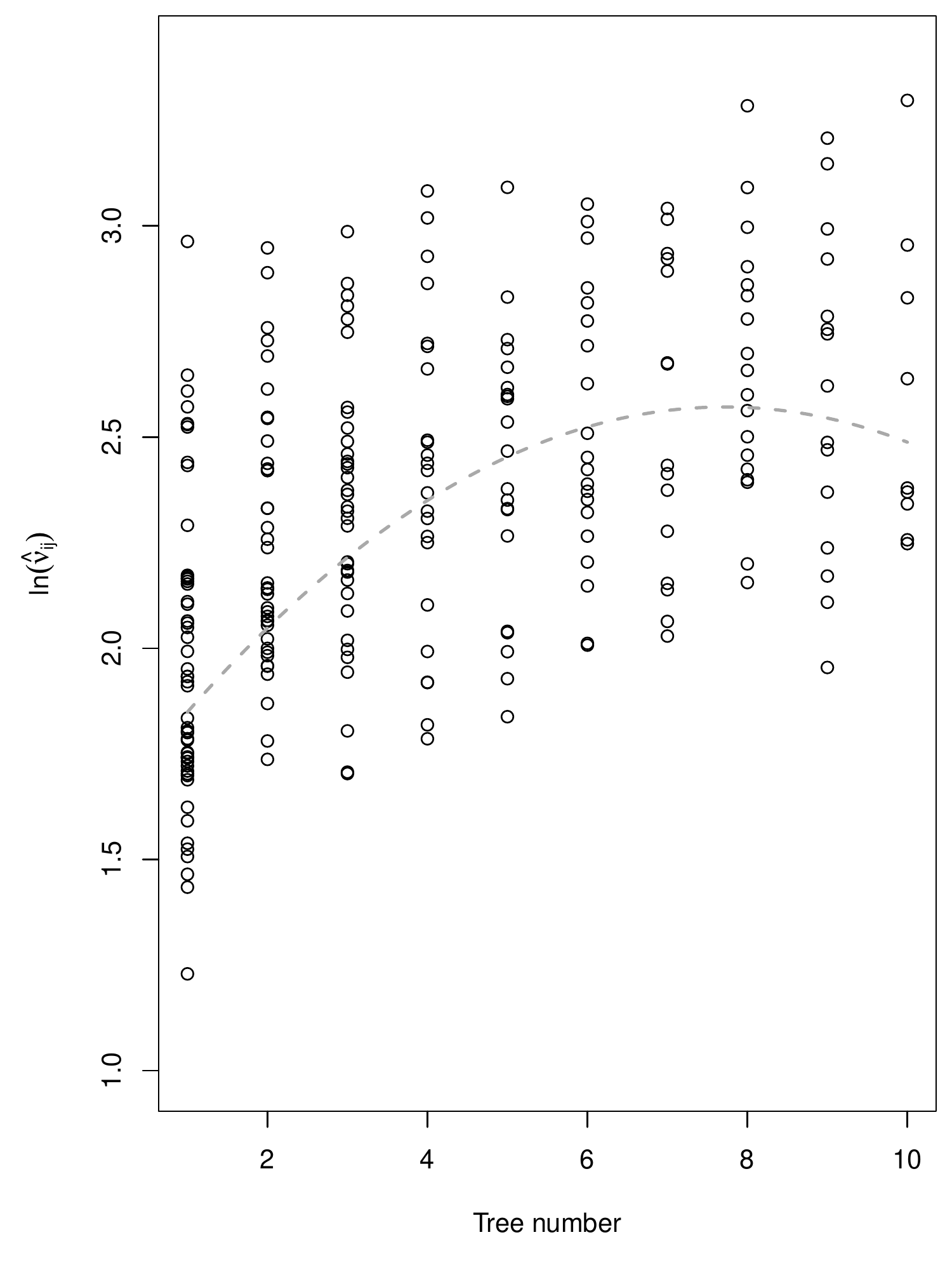}
	\caption[Plot of $\ln\left(\wh\nu_{i(e),j(e);\calD_e}^l\right)$ against tree number.]{Plot of logarithmized estimated degree of freedom parameters $\ln\left(\wh\nu_{i(e),j(e);\calD_e}^l\right)$, for edges $e\in\edge_l$ with Student-$t$ copulas, against the respective tree number $\iN{l}{10}$. The curve given by the model specification \eqref{eq:nurep} using the parameters $\wh\bbeta_\nu^{\text{SV}}$ estimated in Section \ref{sec:app:SVfit} is indicated as a dashed line.}
	\label{fig:treeparameters}
\end{figure}

Figure \ref{fig:treeparameters} plots the logarithmized estimated degrees of freedom parameters $\ln\left(\wh\nu_{i(e),j(e);\calD_e}^l\right)$ of the Student-$t$ copulas which occur in the R-vine against the respective tree number $l$. We discover a quadratic trend (dashed gray line) with regard to the tree number. This finding will be used to model the second copula parameters $\nu_{i(e),j(e);\calD_e}^l$ jointly for all trees $\iN{l}{10}$.

It remains to study the relationships between the first copula parameters $\te_{i(e),j(e);\calD_e}^l$ and the corresponding distances $d_{i(e),j(e)}$ respectively elevation differences $e_{i(e),j(e)}$, distinguishing which tree $l\leq10$ the edge $e$ stems from. We know that there exist relationships
\beqq\label{eq:gtheta}
	\te_{i(e),j(e);\calD_e}^l=g_{\te}\left(\tau_{i(e),j(e);\calD_e}^l;b_{i(e),j(e);\calD_e}\right),
\eeqq
between the copula parameters $\te_{i(e),j(e);\calD_e}^l$ and the Kendall's $\tau$'s $\tau_{i(e),j(e);\calD_e}^l$, depending on the copula family $b_{i(e),j(e);\calD_e}$. Hence we need to investigate possible relationships between the Fisher z-transformed Kendall's $\tau$'s and the distances and elevation differences, separately for each tree.
A similar modeling approach was already followed by \citet[][]{graeler11}.

For the purpose of the tree-wise analysis we define average distances and elevations
\begin{align*}
	\ov{d_{i(e),\calD_e}}	&	\coloneqq \frac{1}{l-1}\sum_{k\in\calD_e}d_{i(e),k},	\quad	\ov{d_{j(e),\calD_e}}	\coloneqq \frac{1}{l-1}\sum_{k\in\calD_e}d_{j(e),k},	\\
	\ov{e_{i(e),\calD_e}}	&	\coloneqq \frac{1}{l-1}\sum_{k\in\calD_e}e_{i(e),k},	\quad	\ov{e_{j(e),\calD_e}}	\coloneqq \frac{1}{l-1}\sum_{k\in\calD_e}e_{j(e),k},
\end{align*}
for all edges $e\in\edge_l$ of trees $\tree_l$ with $l>1$, where the conditioning set $\calD_e$ is non-empty,
and consider them as further potential predictors in our models.

For details on the tree-wise analysis we refer the reader to Chapter 5 of \citet{erhardt13}. The overall picture which we obtain from this analysis is, that in general the distance based predictors capture more dependence information than the elevation based ones and that the direct unconditioned distances have the greatest ability to model Kendall's $\tau$ properly.


\subsection{Model Formulation and Selection}\label{sec:mod:form}

Our preliminary analyses suggest first copula parameter model specifications of the form
\beqq\label{eq:distrep}
	\te_{i(e),j(e);\calD_e}^l \coloneqq \itau\left[\Fzi \left\{h_{l}(e|\bbeta_{l})\right\};b_{i(e),j(e);\calD_e}\right],	\quad e\in\edge_l, \quad \iN{l}{10}.
\eeqq
The inclusion of different combinations of the available spatial predictors $d_{i(e),j(e)}$, $e_{i(e),j(e)}$, $\ov{d_{i(e),\calD_{e}}}$, $\ov{e_{i(e),\calD_{e}}}$, $\ov{d_{j(e),\calD_{e}}}$ and $\ov{e_{j(e),\calD_{e}}}$ into the model is controlled by the model function
$h_{l}\left(e|\bbeta_{l}\right)$, $e\in\edge_l$, $l=1,\ldots,10$,
which is linear in the logarithmized predictors.

A tree-wise comparison of different model specifications in Tables 5.4-5.7 and Figure 5.3.1 in \citet{erhardt13} led to the selection of a model, which includes all available distance based predictors.
The investigations showed, that an additional inclusion of the elevation based predictors wouldn't lead to a significant improvement in terms of explanatory power.

The model function $h_{l}(e|\bbeta_{l})$ of the distance model specification is defined tree-wise. For the first tree the model function is defined as
\beqq\label{eq:h1}
	h_{1}\left(e|\bbeta_{1}\right) \coloneqq \beta_{1,0} + \beta_{1,1}\lnf{d_{i(e),j(e)}}, \quad e\in\edge_1,
\eeqq
with $\bbeta_{1} = \left(\beta_{1,0}, \beta_{1,1}\right)\T \in \R^{2}$.
For all trees $l\geq2$ the model function is given as
\beqq\label{eq:hl}
	h_{l}\left(e|\bbeta_{l}\right)	\coloneqq \beta_{l,0} + \beta_{l,1}\lnf{d_{i(e),j(e)}}	+ \beta_{l,2}\lnf{\ov{d_{i(e),\calD_{e}}}}	+ \beta_{l,3}\lnf{\ov{d_{j(e),\calD_{e}}}}, \quad e\in\edge_l, \quad l=2,\ldots,10,
\eeqq
with parameters $\bbeta_{l}	\coloneqq \left(\beta_{l,0}, \beta_{l,1}, \beta_{l,2}, \beta_{l,3}\right)\T \in \R^{4}$.
We summarize the parameters of the distance model specification as
$\bbeta_{\text{dist}}^{\text{SV}} \coloneqq \left(\bbeta_{1}\T,\ldots,\bbeta_{10}\T\right)\T \in \R^{38}$.

Moreover the investigations in \citet{erhardt13} and Figure \ref{fig:treeparameters} suggested the polynomial model specification for the second copula parameters, which is given by
\beqq\label{eq:nurep}
	\nu_{i(e),j(e);\calD_e}^l	\coloneqq \exp\left\{\beta_{0}^{\nu} + \beta_{1}^{\nu}l + \beta_{2}^{\nu}l^2\right\}, \quad e\in\edge_l, \, \iN{l}{10},
\eeqq
and we define $\bbeta_\nu^{\text{SV}}	\coloneqq \left(\beta_{0}^{\nu}, \beta_{1}^{\nu}, \beta_{2}^{\nu}\right)\T \in \R^{3}$.

\subsection{Model Fit}\label{sec:app:SVfit}

To enable maximum-likelihood estimation, we have to specify the likelihood corresponding to the selected model. Moreover the copula family specification of the truncated R-vine under consideration has to be weakened in terms of family rotation, since the parameters $\te_{i(e),j(e);\calD_e}^l$ may change their sign during the numerical optimization procedure. The final parameter estimates will determine the rotation of the corresponding families. Using the model specification \eqref{eq:nurep} for the degrees of freedom $\nu_{i(e),j(e);\calD_e}^l$ and the model specification \eqref{eq:distrep} for $\te_{i(e),j(e);\calD_e}^l$, the usual R-vine likelihood changes to
\beqo
	\Lik{\text{SV}}{\bbeta_{\text{dist}}^{\text{SV}}, \bbeta_\nu^{\text{SV}}}{\bu^1,\ldots,\bu^d}	= \prod_{t=1}^N\prod_{l=1}^{10}\prod_{e\in\edge_l} \cop{i(e),j(e);\calD_e}{\wt{u}^{i(e)}_t ,\wt{u}^{j(e)}_t\sepi\te_{i(e),j(e);\calD_e}^l,\nu_{i(e),j(e);\calD_e}^l},
\eeqo
where the transformed variables are calculated according to	$\wt{u}^{i(e)}_t	\coloneqq C_{i(e)|\calD_e}(u^{i(e)}_t\sep\bu^{\calD_e}_t)$,	and	$\wt{u}^{j(e)}_t	\coloneqq C_{j(e)|\calD_e}(u^{j(e)}_t\sep\bu^{\calD_e}_t)$, with	$\bu^{\calD_e}_t	\coloneqq \left\{u_t^s: s\in\calD_e\right\}$.
Numerical maximization of the log-likelihood
$\ll{\text{SV}}{\bbeta_{\text{dist}}^{\text{SV}}, \bbeta_\nu^{\text{SV}}}{\ldots} = \ln\Lik{\text{SV}}{\bbeta_{\text{dist}}^{\text{SV}}, \bbeta_\nu^{\text{SV}}}{\ldots}$
yields the maximum-likelihood estimates (mle) $\wh\bbeta_{\text{mle}}^{\text{SV}} = \left({\wh\bbeta_{\text{dist}}^{\text{SV}}}, {\wh\bbeta_\nu^{\text{SV}}}\right)\T \in \R^{41}$. For the results of the estimation procedure and the selection of suitable starting values we refer to Subsection 5.3.2 of \citet{erhardt13}.

Finally we provide an illustration of the dependencies modeled by the spatial R-vine model. Figure \ref{fig:SVGermany} shows all $54$ observation stations on which the spatial R-vine model is fitted and all edges that occur in the ten trees of the fitted R-vine model. The magnitude of association between station pairs is indicated through edge width and edge color. The thicker and darker the edges are, the higher is the respective estimated association. The resulting network gives an impression for which pairs dependencies are modeled and how strong they are.
One clearly observes that the strongest dependencies are already captured in tree one, which is silhouetted against all other trees.

The Student-$t$ copula degrees of freedom resulting from our estimation are visualized in Figure \ref{fig:treeparameters} (dashed gray line). We conclude from the plot, that our model yields strong tail dependencies in the first trees, which get weaker with increasing tree number. The degrees of freedom stay about the same for the trees $\tree_6,\ldots,\tree_{10}$.

\subsection{Prediction}\label{sec:app:SVpred}

Now that we have selected an adequate spatial R-vine model for the mean temperature data and the respective model parameters are estimated, we aim to predict mean temperatures at new locations based on the model fit. In order to be able to validate the outcome of these predictions, we predict for the locations indicated by the validation data introduced in Section \ref{sec:margins} (see Table \ref{tab:scores} for details).

\paragraph{Methodology}\label{sec:SVpred:meth}

Since our spatial R-vine model is constructed based on copula data, predictions from this model will also be on a copula data level. Thus a back transformation to the original level of mean temperatures is needed, which is based on the marginal models presented in Section \ref{sec:margins}. For details on how this technical back transformation is conducted, we refer to Appendix A.

To predict mean temperatures respectively the corresponding copula data $u_t^s$ at a new location $s$ for an arbitrary point in time $t$, we need to specify the conditional distribution $C_{s|1,\ldots,d}(u_t^s|u_t^1,\ldots,u_t^d)$ of the variable $u_t^s$ conditioned on $u_t^1,\ldots,u_t^d$ constituting the copula data at the point in time $t$ given by the training data set on which the spatial R-vine model is built. The spatial R-vine model specifies the joint distribution of $u_t^1,\ldots,u_t^d$, as an R-vine distribution. Therefore, access to the conditional distribution $C_{s|1,\ldots,d}(u_t^s|u_t^1,\ldots,u_t^d)$ can be achieved by extending the underlying spatial R-vine by one further vertex $s$.

If one wants to preserve the structure of the underlying R-vine, one has to add the new variable as a leaf to the first R-vine tree. To do so we estimate the Kendall's $\tau$'s $\tau_{i(e_1),j(e_1)}$ for all $d$ edges $e_1=\{i(e_1),j(e_1)\}=\{r,s\}$, $\iN{r}{d}$, which may be added,
by
\beqo
		\wh\tau_{i(e_1),j(e_1);\calD_{e_1}}	\coloneqq \wh\tau_{i(e_1),j(e_1)} = \wh\tau_{r,s}	= \Fzi\left\{h_{1}\left(e_1=\{r,s\}|\wh\bbeta_{1}\right)\right\}.
\eeqo
Here the conditioning set $\calD_{e_1}$ is the empty set, $h_{1}$ is the model function defined in \eqref{eq:h1} and $g_z$ is given by \eqref{eq:fisher}. The edge $e_1^*$ which yields the biggest Kendall's $\tau$ estimate is selected to extend the first R-vine tree. For this edge a copula family $b_{i(e_1^*),j(e_1^*)}$ has to be selected. We select the copula family which occurs most often in the original R-vine, however other selection criteria might be chosen. The corresponding first copula parameter $\wh\theta_{i(e_1^*),j(e_1^*)}^1 = \wh\theta_{i(e_1^*),j(e_1^*);\calD_{e_1^*}}^1$ is estimated as
\beqq\label{eq:1cp}
		\wh\theta_{i(e_1^*),j(e_1^*);\calD_{e_1^*}}^1 = \itau\left\{\wh\tau_{i(e_1^*),j(e_1^*);\calD_{e_1^*}}\sepi b_{i(e_1^*),j(e_1^*)}\right\}
\eeqq
using \eqref{eq:gtheta}. If needed the second copula parameter $\wh\nu_{i(e_1^*),j(e_1^*);\calD_{e_1^*}}^1 = \wh\nu_{i(e_1^*),j(e_1^*)}^1$ is estimated as
\beqq\label{eq:2cp}
		\wh\nu_{i(e_1^*),j(e_1^*);\calD_{e_1^*}}^1 = h_{\nu}\left(e_1^*,1|\wh\bbeta_{\nu}^{\text{SV}}\right),
\eeqq
where the function $h_{\nu}\left(e,l|\bbeta_{\nu}^{\text{SV}}\right)$, which depends on the respective edge $e$ and tree number $l$ and is parametrized by $\bbeta_{\nu}^{\text{SV}}$, represents the model specification for the second copula parameters (see Equation \eqref{eq:nurep}). 

The rest of the R-vine is extended tree-wise starting from tree number two. For each tree $l$ we have to ensure that the proximity condition is fulfilled after a new edge $e_l$ has been added. For all edges $e_l$ with $j(e_l)=s$ and $\calD_{e_l}=\calD_{e_{l-1}^*}\cup i(e_{l-1}^*)$ which fulfill the proximity condition, we estimate the corresponding Kendall's $\tau$'s using \eqref{eq:hl} by
\beqo
		\wh\tau_{i(e_l),j(e_l);\calD_{e_l}} = \Fzi\left\{h_{l}\left(e_l|\wh\bbeta_{l}\right)\right\}.
\eeqo
Again the edge $e_l^*$ with the biggest Kendall's $\tau$ estimate is selected and included into the R-vine and a copula family $b_{i(e_l^*),j(e_l^*);\calD_{e_l^*}}$ has to be selected. The corresponding parameters $\theta_{i(e_l^*),j(e_l^*);\calD_{e_l^*}}^l$ and $\nu_{i(e_l^*),j(e_l^*);\calD_{e_l^*}}^l$ have to be estimated in analogy to \eqref{eq:1cp} and \eqref{eq:2cp}, respectively.

For trees exceeding the truncation level $k<d$, arbitrary edges which fulfill the proximity condition can be chosen. The copulas corresponding to these edges are selected to be independence copulas. Thus, no parameters have to be specified for these copulas.

The above described procedure yields an R-vine copula specification corresponding to the variables $u_t^s, u_t^1,\ldots,u_t^d$ with R-vine distribution $C(u_t^s, u_t^1,\ldots,u_t^d)$. Applying Equation \eqref{eq:joe}, this allows to calculate $C_{s|1,\ldots,d}(u_t^s|u_t^1,\ldots,u_t^d)$ iteratively. Thus we are able to simulate from the predictive distribution $C_{s|1,\ldots,d}(u_t^s|u_t^1,\ldots,u_t^d)$ using the probability integral transform. We simulate $v\sim\vU01$ and set $\check{u}_t^s \coloneqq C_{s|1,\ldots,d}^{-1}(v|u_t^1,\ldots,u_t^d)$ as a simulation of the copula data point at location $s$ at time $t$.

If one transforms the copula data $\check{u}_t^s$ resulting from these simulations back to the level of the originally modeled data $\check{y}_t^s$, one can calculate point predictions $\wh{y}_t^s$ as the mean of the back transformed simulations.

Now we discuss how to obtain the corresponding prediction density. Omitting all arguments, the prediction density $c_{s|1,\ldots,d}$ corresponding to $C_{s|1,\ldots,d}$ can be obtained by decomposing numerator and denominator of $c_{s|1,\ldots,d} = \nicefrac{c_{s,1,\ldots,d}}{c_{1,\ldots,d}}$
according to Equation \eqref{eq:vcdensity} into products of pair copulas. Since the R-vine copula specification corresponding to $c_{s,1,\ldots,d}$ differs from the R-vine copula specification corresponding to $c_{1,\ldots,d}$ only in terms of the additional edges $e_1^*,\ldots,e_{d-1}^*$, and due to the fact that it holds $j(e_l^*)=s$ by construction and that we consider truncations at a certain level $k<d$, we obtain
\beqo
		c_{s|1,\ldots,d}	= \prod_{l=1}^{k} \cop{i(e_l^*),s;\calD_{e_l^*}}{C_{i(e_l^*)|\calD_{e_l^*}},C_{s|\calD_{e_l^*}}\sepi\wh\te_{i(e_l^*),s;\calD_{e_l^*}}^l,\wh\nu_{i(e_l^*),s;\calD_{e_l^*}}^l}.
\eeqo

In our case we perform the above calculations based on the distance model specification \eqref{eq:distrep} and on the model specification \eqref{eq:nurep} for the second copula parameters. Due to our previous investigations on the structure of the R-vine underlying the spatial R-vine model (see Table \ref{tab:SV:struct}) we select a Student-$t$ copula for every edge which is added to the truncated R-vine. The subsequently discussed predictions of the $19$ mean temperature time series
constituting the validation data set are based on $1000$ simulations of each time series.

\paragraph{Results}\label{sec:SVpred:results}

We select the two stations \emph{Grambek} ($67$) and \emph{Arkona} ($56$) as representatives for the forthcoming analysis. The respective predictions are compared in Figure \ref{fig:SVpredMSE}. For the purpose of comparison we plotted the observed values in black and the prediction in gray. Moreover, the corresponding $95\%$ prediction intervals are indicated by the light gray area around the point predictions. Whereas the predictions for \emph{Grambek} are very close to the observed values and the prediction intervals are very narrow, we observe noticeable deviations for \emph{Arkona}. There seems to be more uncertainty in the predictions for \emph{Arkona}, which is reflected in the comparatively broad prediction intervals. This might be due to the special location of \emph{Arkona} on an island in the Baltic Sea, where the temperatures might be exposed to several factors which are not included in our model.
Figure \ref{fig:SVpredresMSE} highlights the prediction errors for the two previously selected stations \emph{Grambek} ($67$) and \emph{Arkona} ($56$). For \emph{Arkona} we observe systematic deviations from zero, which alludes to a misspecification of the seasonality parameters. This may be due to the fact that the latitude of \emph{Arkona} lies outside the latitude range of our training data set.

This first analysis of predictions from our spatial R-vine model yields an impression of the prediction capabilities and limitations of our model. We see a good prediction performance, as long as we predict within the observed modeling framework. However as expected our marginal model is not able to capture the temperature trends of stations which lie outside the range of the training data set.


\section[A Spatial Gaussian Model for Daily Mean Temperatures]{A Spatial Gaussian Model for Daily Mean Temperatures}\label{sec:classical}

For comparison we introduce a \textit{spatial Gaussian model} (SG).

\paragraph[The Model]{The Model}

As before let $\wt Y_t^s$ be a real valued random variable, which represents the (weighted) mean temperature at a location $s$ and a point in time $t$. Let moreover $\wt\bY_t\coloneqq(\wt Y_t^1,\ldots,\wt Y_t^d)\T\in\R^d$ for all $\iN{t}{N}$. Then our spatial Gaussian model is given by
\beqo
	\wt \bY_t = \bmu_t + \bveps_t, \quad \bveps_t\simiid\mvN{d}{\bnull}{\Sigma(\btheta^{\text{SG}})}, \quad \iN{t}{N},
\eeqo
where $\bmu_t\coloneqq(\mu_t^1,\ldots,\mu_t^d)\T\in\R^d$ is a vector of means for all $\iN{t}{N}$ and $\Sigma(\btheta)\in\R^{d\times d}$ is a positive definite covariance matrix depending on some $n_\text{par}^\Sigma$-dimensional parameter vector $\btheta^{\text{SG}}$. The components of the mean vector $\bmu_t$ are modeled analogous to Equation \eqref{eq:jointmargmod}.
The spatial dependencies are determined by the covariance matrix $\Sigma(\btheta^{\text{SG}})=\left\{\Sigma_{i,j}(\btheta^{\text{SG}})\right\}_{i,\iN{j}{d}}$ which in turn is modeled based on a \emph{Gaussian variogram model} \citep[see for example][Chapter 3]{gelfand10}
$\gamma(h \sepi \eta, \varsigma, \rho) \coloneqq \varsigma\left\{1-\exp\left(-\frac{h^2}{\rho^2}\right)\right\} + \eta{\mathds 1}_{(0,\infty)}(h)$.
Then the variance is given as $\sigma^2 = \lim_{h \to \infty} \gamma(h \sepi \eta, \varsigma, \rho) = \eta + \varsigma$ and we model
$\Sigma_{i,j}(\btheta^{\text{SG}}) \coloneqq \sigma^2 - \gamma(d_{i,j} \sepi \eta, \varsigma, \rho)$,
where the parameter vector $\btheta^{\text{SG}}$ consists of the three components $\eta, \varsigma, \rho$. Here $d_{i,j}$ are the pairwise distances between the observation station pairs $(i,j)$, $i,\iN{j}{d}$. By doing so we implicitly make a stationarity assumption.

Comparing the spatial R-vine and the spatial Gaussian model we use the same mean function, however the distribution of the residuals is modeled differently. I.e. in the case of the spatial R-vine model we utilize skew-$t$ marginals and an R-vine copula compared to Gaussian marginals and a Gauss copula for the spatial Gaussian model.

\paragraph[Parameter Estimation]{Parameter Estimation}

Parameters are estimated in two steps. First the mean vectors $\bmu_t$, $\iN{t}{N}$, are estimated using least-squares estimation of the parameter vector $\bbeta$. This is done in the same way as for the marginal model in Subsection \ref{sec:mmpe} and we obtain the same estimates $\wh\bbeta$. Based on these estimates we calculate the residual vectors $\wh\bveps_t \coloneqq \wt\by_t-\wh\bmu_t$. In a second step we perform maximum likelihood estimation of the parameters $\btheta^{\text{SG}}=(\eta, \varsigma, \rho)\T$ using the log-likelihood
\beqq\label{eq:llSG}
	\ll{\text{SG}}{\btheta^{\text{SG}}}{\xn{\wh\bveps}{N}}	= -\frac{N}{2}\ln\left\{(2\pi)^{d}\left|\Sigma(\btheta^{\text{SG}})\right|\right\} -\half\sumtN\wh\bveps_t\T\left\{\Sigma(\btheta^{\text{SG}})\right\}^{-1}\wh\bveps_t.
\eeqq

\paragraph[Prediction]{Prediction}

For the purpose of prediction of mean temperatures at a new location $o$ we assume that the mean temperatures $\wt Y_t^o$, $\iN{t}{N}$ follow the model specified above. Thus we assume that
\beqo
	\begin{pmatrix}
		\wt Y_t^o\\
		\wt\bY_t
	\end{pmatrix}
	=
	\begin{pmatrix}
		\mu_t^o\\
		\bmu_t
	\end{pmatrix}
	+
	\begin{pmatrix}
		\veps_t^o\\
		\bveps_t
	\end{pmatrix}, \quad
	\begin{pmatrix}
		\veps_t^o\\
		\bveps_t
	\end{pmatrix}\simiid\mvN{d+1}{\bnull}{\Sigma^*(\btheta^{\text{SG}})},
\eeqo
where the covariance matrix extends to
\beqo
	\Sigma^*(\btheta)	=
	\left\{\begin{array}{c|c}
		\sigma^2	&	\bsigma_{o}\T	\\
		\hline
		\bsigma_{o}	&	\Sigma(\btheta^{\text{SG}})
	\end{array}\right\} \in\R^{d+1},
\eeqo
with $\bsigma_{o} \coloneqq \{\sigma^2-\gamma(d_{1,o} \sepi \eta, \varsigma, \rho),\ldots, \sigma^2-\gamma(d_{d,o} \sepi \eta, \varsigma, \rho)\}\T\in\R^d$. Using basic results for the conditional distribution of a mulitvariate normal distribution \citep*[see e.g.][Section 3.4]{eaton07}, this yields that
$\veps_t^o|\bveps_t=\wh\bveps_t \simiid\vN{\ov{\mu}(\btheta^{\text{SG}})}{\ov{\Sigma}(\btheta^{\text{SG}})}$, $\iN{t}{N}$,
with $\ov{\mu}(\btheta^{\text{SG}})=\bsigma_{o}\T\left\{\Sigma(\btheta^{\text{SG}})\right\}^{-1}\wh\bveps_t$ and $\ov{\Sigma}(\btheta^{\text{SG}})=\sigma^2 - \bsigma_{o}\T\left\{\Sigma(\btheta^{\text{SG}})\right\}^{-1}\bsigma_{o}$.

We perform the prediction in analogy to our vine copula based model using simulation. We repeatedly simulate $\check{\veps}_t^o$ from $\calN\left\{\ov{\mu}(\wh\btheta^{\text{SG}}),\ov{\Sigma}(\wh\btheta^{\text{SG}})\right\}$ and perform a back transformation to the original data level based on the marginal model to achieve mean temperature simulations $\check{y}_t^o$. For details on this back transformation we refer again to Appendix A.

The results from the predictions for the two selected stations \emph{Grambek} ($67$) and \emph{Arkona} ($56$) are illustrated in Figures \ref{fig:STpredMSE} and \ref{fig:STpredresMSE}.

\section[Model Validation and Comparison]{Model Validation and Comparison}\label{sec:appl}

\begin{figure}[htb]
	\centering
		\includegraphics[width=1.00\textwidth]{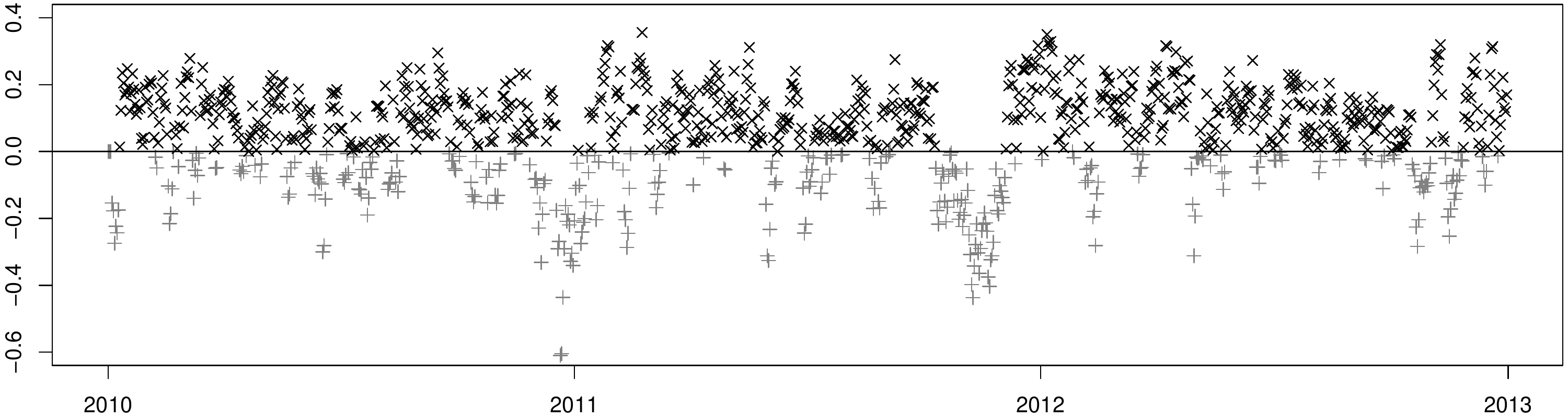}
	\caption[Log-score difference plots of the averaged continuous ranked probability scores comparing the different models (averaged over all $19$ observation stations of the validation data set).]{Log-score difference plots of the averaged continuous ranked probability scores comparing the spatial R-vine model to the corresponding averaged spatial Gaussian model scores (average over all $19$ observation stations of the validation data set). Points in time where the spatial R-vine model has the lower average scores are marked by a black x. On the other hand, points in time where the spatial Gaussian model has the lower average scores are marked by a gray plus sign.}
	\label{fig:AverageScoresOP}
\end{figure}

For the purpose of model comparison we calculate (negatively oriented) continuous ranked probability scores (CRPS) \citep[see][Section 4.2]{gneiting07}. Negatively oriented means that smaller scores, i.e. scores closer to zero indicate a better fit. The scores will allow for an adequate comparative model validation. In the following we consider averaged continuous ranked probability scores (Table \ref{tab:scores}, $\ov{\text{CRPS}}$), percentaged model outperformance (Table \ref{tab:scores}, $\% \left(\text{SV} \succ \text{SG}\right)$) and a new concept called log-score difference plots (Figure \ref{fig:AverageScoresOP}).

\paragraph{Averaged scores}
In order to get a first impression which model provides better predictions, we compare the averaged continuous ranked probability scores ($\ov{\text{CRPS}}$) in Table \ref{tab:scores}, where we average over time. Moreover the overall averages are given in the last row of Table \ref{tab:scores}. Since scores close to zero are preferred, the overall consideration of the averaged scores in Table \ref{tab:scores} yields, that we prefer the spatial R-vine model.

\begin{table}[htb]
\centering
\caption[Comparison of averaged CRPS and percentaged outperformance.]{Comparison of the averaged CRPS of the spatial R-vine model (SV) and the spatial Gaussian model (SG) and percentaged outperformance ($\% \left(\text{SV} \succ \text{SG}\right)$) in terms of CRPS over the period $01/01/2010-12/31/2012$ for the observation stations of the validation data set. Here we define $\% \left(\text{SV} \succ \text{SG}\right)$ as the share of the points in time for which the spatial R-vine model is preferred over the spatial Gaussian model in terms of CRPS.} 
\label{tab:scores}
\begin{tabular}{rllccc}
	\hline
	\hline
		&  & short & \multicolumn{2}{c}{$\ov{\text{CRPS}}$}  &		\\ 
	s	& name & name & SV & SG &	$\% \left(\text{SV} \succ \text{SG}\right)$	\\ 
  \hline
	55 & Alfeld & \texttt{alfe} & 3.20 & 2.59 & 0.18 \\ 
  56 & \textbf{Arkona} & \texttt{arko} & 3.11 & 3.44 & 0.72 \\ 
  57 & Arnsberg-Neheim & \texttt{arns} & 2.25 & 2.61 & 0.79 \\ 
  58 & Augsburg & \texttt{augs} & 2.73 & 2.57 & 0.48 \\ 
  59 & Blankenrath & \texttt{blan} & 3.00 & 2.64 & 0.33 \\ 
  60 & Borkum-Flugplatz & \texttt{bork} & 2.32 & 3.22 & 0.93 \\ 
  61 & Bremerv\"orde & \texttt{bvoe} & 2.32 & 2.59 & 0.73 \\ 
  62 & Buchen, Kr. Neckar-Odenwald & \texttt{buch} & 2.54 & 2.60 & 0.61 \\ 
  63 & Coschen & \texttt{cosc} & 2.65 & 2.84 & 0.68 \\ 
  64 & Ebrach & \texttt{ebra} & 2.33 & 2.57 & 0.73 \\ 
  65 & Ellwangen-Rindelbach & \texttt{ellw} & 3.22 & 2.59 & 0.20 \\ 
  66 & Falkenberg, Kr.Rottal-Inn & \texttt{falk} & 2.64 & 2.61 & 0.57 \\ 
  67 & \textbf{Grambek} & \texttt{gram} & 1.84 & 2.54 & 0.93 \\ 
  68 & Gr\"unow & \texttt{grue} & 1.98 & 2.65 & 0.91 \\ 
  69 & L\"uchow & \texttt{luec} & 2.21 & 2.59 & 0.79 \\ 
  70 & M\"ullheim & \texttt{muel} & 2.22 & 3.03 & 0.91 \\ 
  71 & Oberharz am Brocken-Stiege & \texttt{ohrz} & 3.66 & 2.59 & 0.06 \\ 
  72 & Rahden-Varl & \texttt{rahd} & 2.36 & 2.60 & 0.73 \\ 
  73 & Wiesenburg & \texttt{wies} & 2.72 & 2.60 & 0.45 \\ 
   \hline
  mean &  &  &  \textbf{2.59} & 2.71 & \textbf{0.62} \\ 
   \hline\end{tabular}
\end{table}

\paragraph{Percentaged outperformance}
Furthermore Table \ref{tab:scores} compares both spatial models using percentaged outperformance. For all stations in the validation data set we count for how many points in time the spatial R-vine model yields a lower score than the spatial Gaussian model. For more than two thirds of the stations of the validation data set and for a share of $62\%$ of all temperature predictions under consideration we observe an outperformance of the spatial R-vine model.

\paragraph{Log-score difference plots}
It is possible that the model outperformance depends on the time, i.e. there may be time intervals in which one model yields better results than the other. In order to be able to detect such kinds of time dependencies, we consider Figure \ref{fig:AverageScoresOP}. We call this figure log-score difference plot, since it shows the difference of the logarithmized (negatively oriented) scores of two models against the respective points in time. More precisely the figure depicts the log-score difference plots of the continuous ranked probability scores averaged over all $19$ observation stations of the validation data set. From the plot we see, that there are time intervals towards the end of each year, where the spatial Gaussian model consequently yields lower scores than the spatial R-vine model, while the opposite is true for the remainder of the year.

\section{Discussion}\label{sec:discuss}

An extensive analysis of an ordinary (truncated) R-vine copula fitted to the training data led to a new model for spatial dependencies, the spatial R-vine model. The investigation of the relationship between the Kendall's $\tau$'s occurring in the R-vine copula and the distances and elevation differences which can be associated to these Kendall's $\tau$'s proposed different kinds of tree-wise model specifications for the first pair copula parameters. We found that the explanatory power of the elevation differences is comparatively small, whereas the station distances are able to explain the respective dependencies to a large extent. Therefore we selected a model accounting for all distances between the observation stations, which are associated to the respective bivariate copulas of the R-vine copula specification.

Moreover a model specification for the second copula parameters affecting the large share of Student-$t$ copulas was applied to reduce the necessary number of parameters further. This resulted in the modeling of strong tail dependencies in the lower trees, which distinguishes our spatial R-vine model from classical Gaussian approaches to model such kind of data.

All in all the selected model specifications led to a distinct reduction in the number of parameters. In the case of our example data set, the $733$ parameters needed in the original truncated R-vine copula model could be replaced by $41$ parameters in the spatial R-vine model. This reduction is also mirrored in the computation time of the full maximum likelihood estimation for both models. Whereas the estimation for the truncated R-vine took about $3.7$ days, this time could be reduced to $18$ hours for our spatial R-vine model.

For the purpose of comparison we introduced a spatial Gaussian model, which requires only three parameters. Our aim was it to show that our new approach yields better predictions, which will justify a longer computation time. A validation of the prediction results from both models in terms of continuous ranked probability scores (CRPS) yielded reasonable accuracy of our predictions, as long as the location from which we aimed to predict lay within the range of the training data. Comparison of the continuous ranked probability scores over time revealed a time dependency of the relative prediction performance of both models. The overall consideration of the scores showed an outperformance in $62\%$ of all considered points in time. Transformation of the maximum log-likelihood of the truncated and the spatial R-vine model to the residual level on which the spatial Gaussian model is built, allows an comparison. For the truncated R-vine model, the spatial R-vine model and the spatial Gaussian model we obtain maximum log-likelihoods (residual level) of $-42515.22$, $-46231.80$ and $-49095.58$, respectively. The corresponding AIC and BIC values can be calculated from the respective numbers of parameters $33+733=766$, $33+41=74$ and $3$. AIC, BIC and the log-likelihoods result in the same ranking of the models. Their values show a clear preference of our spatial R-vine model over the spatial Gaussian model.

With regard to future work on the topic of vine copula based models for spatial dependencies an application of our modeling approach to other types of data sets is desirable, which requires the development of appropriate marginal models. Especially an investigation of data sets where asymmetries of bivariate dependencies are observed should stand in the focus of further research. Moreover an improvement respectively extension of our model by the inclusion of further covariates could be investigated. Covariates of interest may be microclimatic variates like \emph{urban/rural area}, \emph{closeness to body of water} or \emph{wind force}.

\section*{Acknowledgements}

The first author likes to thank the TUM Graduate School's Graduate Center International Graduate School of Science and Engineering (IGSSE) for support. The numerical computations were performed on a Linux cluster supported by DFG grant INST 95/919-1 FUGG. \vspace*{-8pt}

%

\bibliographystyle{chicago}	
\bibliography{literature}

\appendix

\renewcommand \thesection {Appendix \Alph{section}}
\setcounter{figure}{0}
\renewcommand\thefigure{\Alph{section}.\arabic{figure}}

\section{Back Transformation}\label{app}

Since the simulations from the spatial R-vine model are on copula data level and the simulations from the spatial Gaussian model are on residual level, a back transformation to the original level of mean temperatures is needed. We build our back transformation procedure based on the marginal model developed in Section 4 of the main article.

For the purpose of the back transformation we need estimates $\wh\beta_0(s)$, $\wh\beta_{\text{s}}(s)$, $\wh\beta_{\text{c}}(s)$, $\wh\gamma_1(s)$, $\wh\gamma_2(s)$, $\wh\gamma_3(s)$, $\wh\xi(s)$, $\wh\omega(s)$, $\wh\alpha(s)$ and $\wh\nu(s)$ of the spatially varying marginal model parameters for the location $s$ under consideration. These estimates are calculated according to Section 4.2, based on the respective parameter estimates. Only $\wh\xi(s)$ is set to $\wh\xi(s) \coloneqq -\wh\omega(s)\wh\mu(s)$, where $\wh\mu(s)$ is calculated according to
\beqo
	\mu\coloneqq\sqrt{\frac{\alpha^2\nu}{(1+\alpha^2)\pi}} \frac{\Gamma\left(\frac{\nu-1}{2}\right)}{\Gamma\left(\frac{\nu}{2}\right)},
\eeqo
to ensure the zero mean condition for the marginal model errors ($\E(X)=\xi+\omega\mu=0$ for $X\sim\vst{\xi}{\omega}{\alpha}{\nu}$).

The first step of the back transformation is the transformation of the copula data\footnote{Remember, that due to the inclusion of the three autoregression components in the marginal model there are no copula data available for the first three points in time.} $\check u_{4}^s,\ldots,\check u_{N}^s$ to the marginal model residuals $\check\veps_{4}^s,\ldots,\check\veps_{N}^s$. This is achieved by means of the quantile function $F_{\text{skew-}t}^{-1}$ of the skew-$t$ distribution with parameters $\wh\xi(s)$, $\wh\omega(s)$, $\wh\alpha(s)$ and $\wh\nu(s)$, i.e. we calculate
\beqo
	\check\veps_t^s = \qst{\wh\xi(s)}{\wh\omega(s)}{\wh\alpha(s)}{\wh\nu(s)}{\check u_t^s}, \quad t=4,\ldots,N.
\eeqo

The only difficulty which arises for the back transformation is due to the autoregression components in the marginal model. In order to obtain the mean temperature series $\check y_{4}^s,\ldots,\check y_{N}^s$, we have to determine meaningful start values for this time series for $t=1,2,3$. We proceed by predicting $\check y_{1}^s$, $\check y_{2}^s$ and $\check y_{3}^s$ based on three linear models of the form
\beqo
	Y_t^s = \theta_0 + \theta_\text{el} x_{\text{el},s} + \theta_\text{lo} x_{\text{lo},s} + \theta_\text{la} x_{\text{la},s} + \veps_t^s, \quad \veps_t^s \simiid \vN{0}{\sigma^2},
\eeqo
for $t=1,2,3$, where the respective parameters are estimated based on the training data ($\iN{s}{54}$). We have to divide the initial predictions $\check y_{1}^s$, $\check y_{2}^s$ and $\check y_{3}^s$ by their respective weights $\wh w_1$, $\wh w_2$ and $\wh w_3$ which results in $\check{\wt y}_{1}^s$, $\check{\wt y}_{2}^s$ and $\check{\wt y}_{3}^s$.

In a last step the weighted mean temperatures $\check{\wt y}_{4}^s,\ldots,\check{\wt y}_{N}^s$ can be calculated as
\beqo
		\check{\wt y}_{t}^s = \wh\beta_0^r(s) + \wh\beta_{\text{s}}(s) \sin\left(\frac{2\pi t}{365.25}\right) + \wh\beta_{\text{c}}(s) \cos\left(\frac{2\pi t}{365.25}\right) + \wh\gamma_1^r(s) \check{\wt y}_{t-1}^s + \wh\gamma_2^r(s) \check{\wt y}_{t-2}^s + \wh\gamma_3^r(s) \check{\wt y}_{t-3}^s + \check\veps_t^s,
\eeqo
where $t=4,\ldots,N$ and $\wh\beta_0(s)$, $\wh\beta_{\text{s}}(s)$, $\wh\beta_{\text{c}}(s)$, $\wh\gamma_1(s)$, $\wh\gamma_2(s)$, $\wh\gamma_3(s)$ are the aggregated parameter estimates for location $s$. Finally we obtain the unweighted mean temperatures as
\beqo
		\check y_{t}^s = \check{\wt y}_{t}^s\sqrt{\wh w_t}, \quad t=1,\ldots,N.
\eeqo

\section{Supplementary Figures}\label{app2}

\begin{figure}[htb]
	\centerline{
		\entrymodifiers={++[o][F]}
		\xymatrix @+2pc {
		*{\tree_1:}	& 1 \ar@{-}[r]^{12}				& 2 \ar@{-}[r]^{23}					& 3 \ar@{-}[r]^{34}	& 4 \\
		*\txt{}	& *\txt{}											& *\txt{}										& 5 \ar@{-}[u]_{35}	& *\txt{}	\\
		*{\tree_2:}	& 12 \ar@{-}[r]^{13|2}		& 23 \ar@{-}[r]^{24|3}			& 34								& *\txt{} \\
		*\txt{}	& *\txt{}											& 35 \ar@{-}[u]_{25|3}			& *\txt{}						& *\txt{}	\\
		*{\tree_3:}	& 13|2 \ar@{-}[r]^{14|23}	& 24|3 \ar@{-}[r]^{45|23}		& 25|3							& *\txt{} \\
		*{\tree_4:}	& *\txt{}									& 14|23 \ar@{-}[r]^{15|234}	& 45|23							& *\txt{} \\
		}
	}
	\caption{Example for an R-vine tree structure.}
	\label{fig:rvine}
\end{figure}
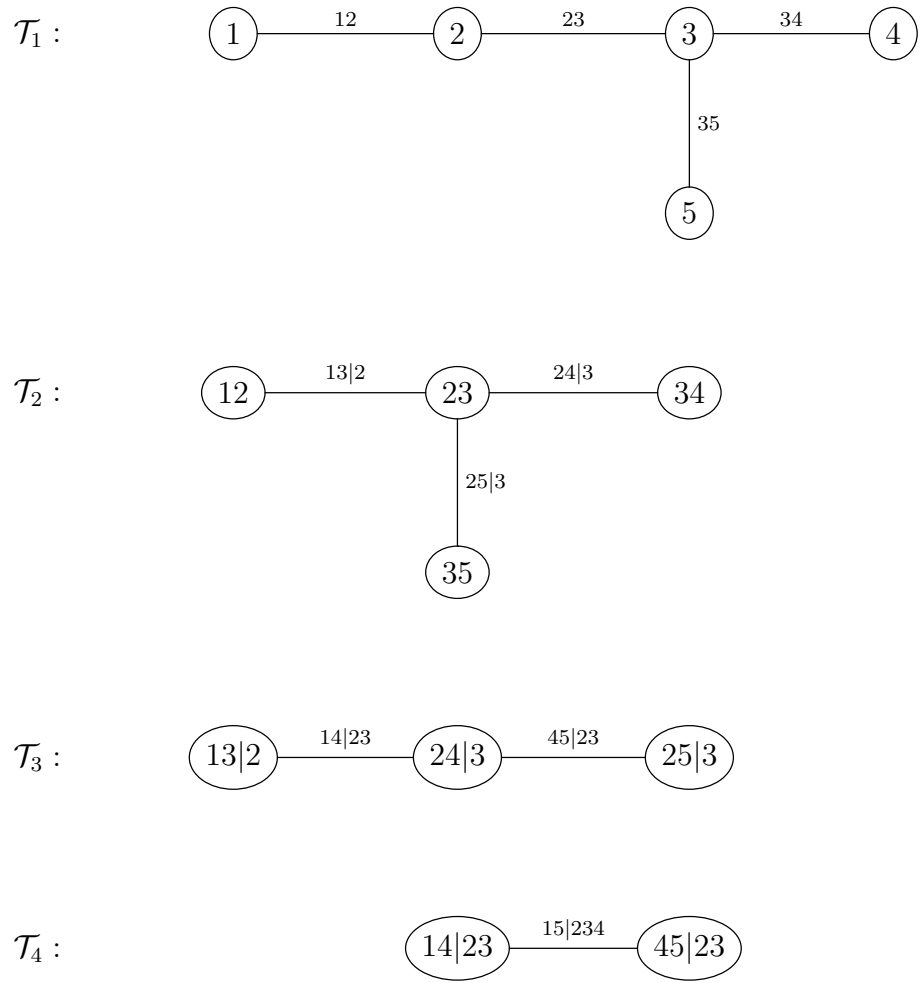

\begin{figure}[htb]
	\centering
		\includegraphics[width=0.95\textwidth]{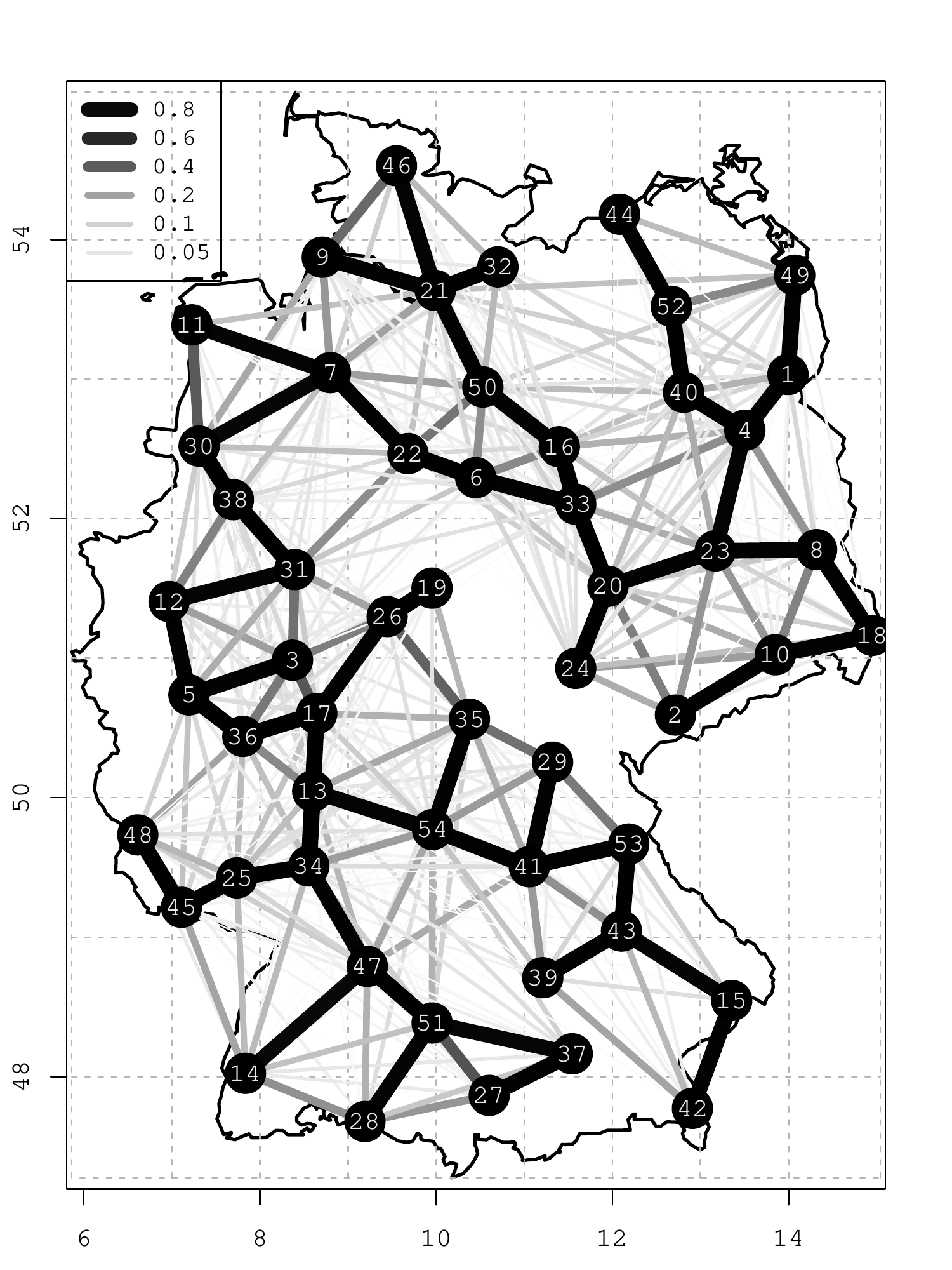}
	\caption[Visualization of the dependence structure in the estimated spatial R-vine model.]{Visualization of the dependence structure in the estimated spatial R-vine model. The edges of all ten trees of the truncated R-vine are depicted. The thicker and darker the edges are, the higher is the respective association.}
	\label{fig:SVGermany}
\end{figure}

\begin{sidewaysfigure}
	\centering
		\includegraphics[width=1.0\textwidth]{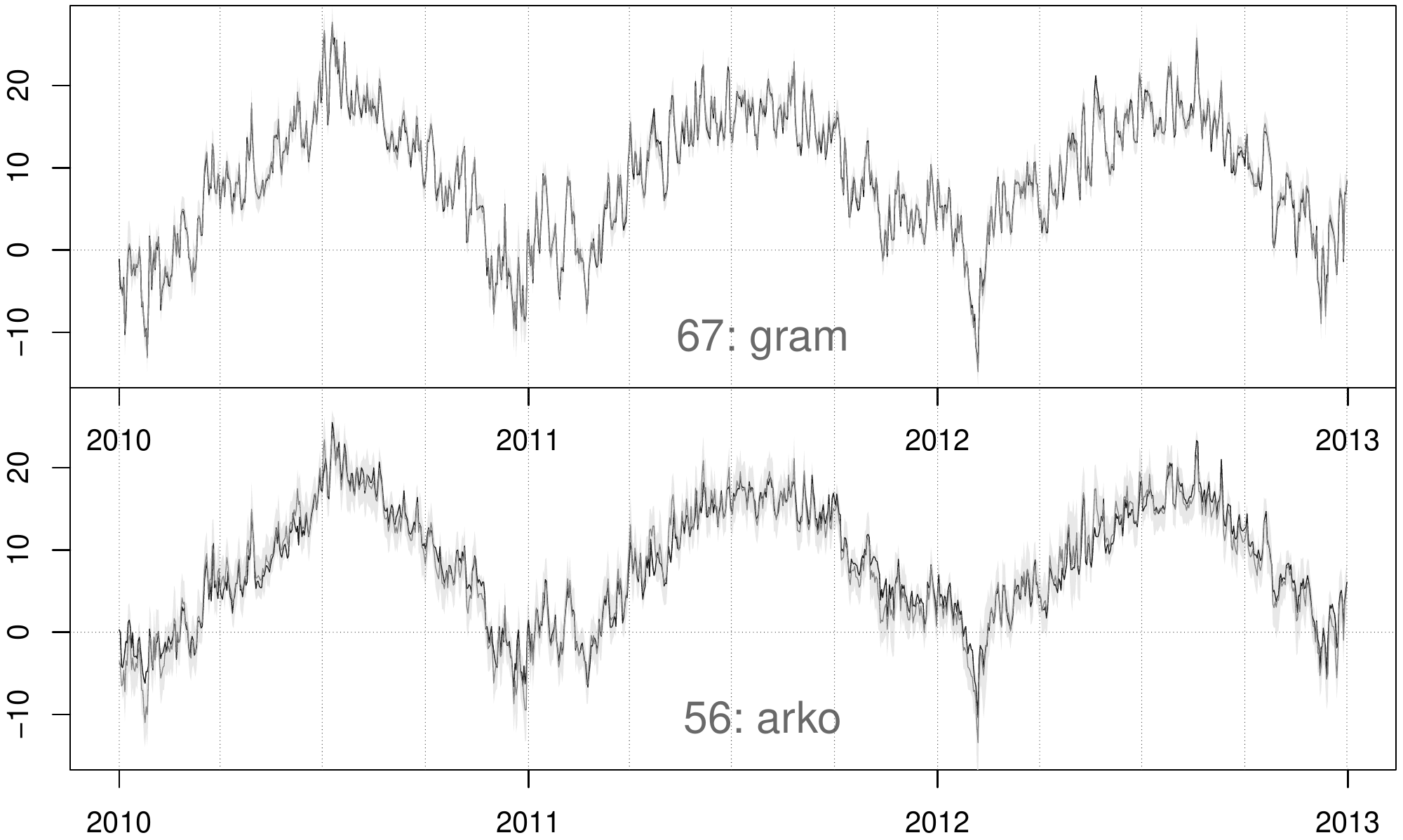}
	\caption[Prediction of the mean temperatures for the observation stations \emph{Grambek} ($67$) and \emph{Arkona} ($56$) based on the spatial R-vine model.]{Prediction of the mean temperatures for the observation stations \emph{Grambek} ($67$) and \emph{Arkona} ($56$) for the period 01/01/2010-12/31/2012 based on the spatial R-vine model. black line: observed values. dark gray line: prediction. light gray area: $95\%$ prediction intervals.}
	\label{fig:SVpredMSE}
\end{sidewaysfigure}

\begin{sidewaysfigure}
	\centering
		\includegraphics[width=1.0\textwidth]{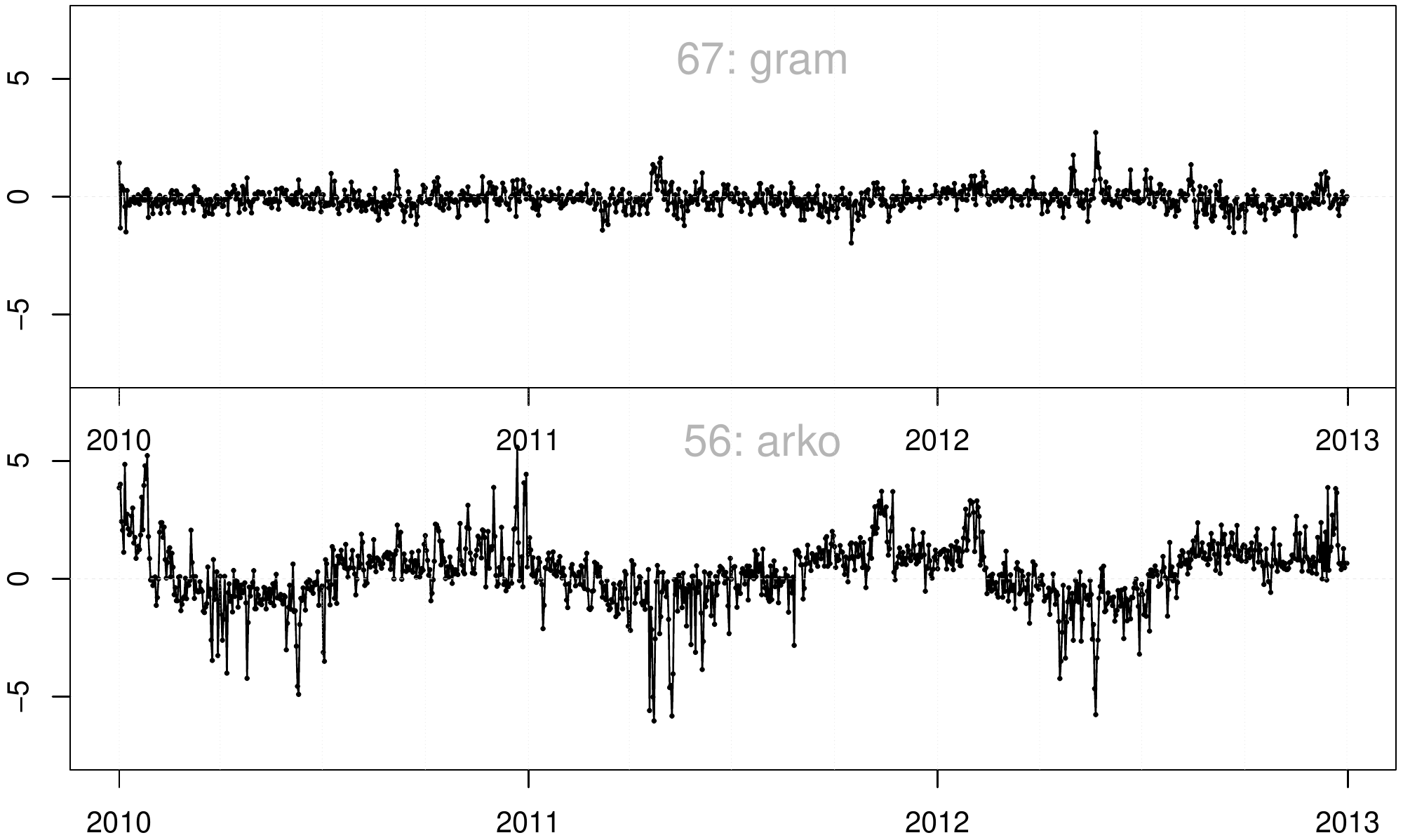}
	\caption[Prediction errors of the predictions for the observation stations \emph{Grambek} ($67$) and \emph{Arkona} ($56$) based on the spatial R-vine model.]{Prediction errors of the predictions for the observation stations \emph{Grambek} ($67$) and \emph{Arkona} ($56$) for the period 01/01/2010-12/31/2012 based on the spatial R-vine model.}
	\label{fig:SVpredresMSE}
\end{sidewaysfigure}

\begin{sidewaysfigure}
	\centering
		\includegraphics[width=1.0\textwidth]{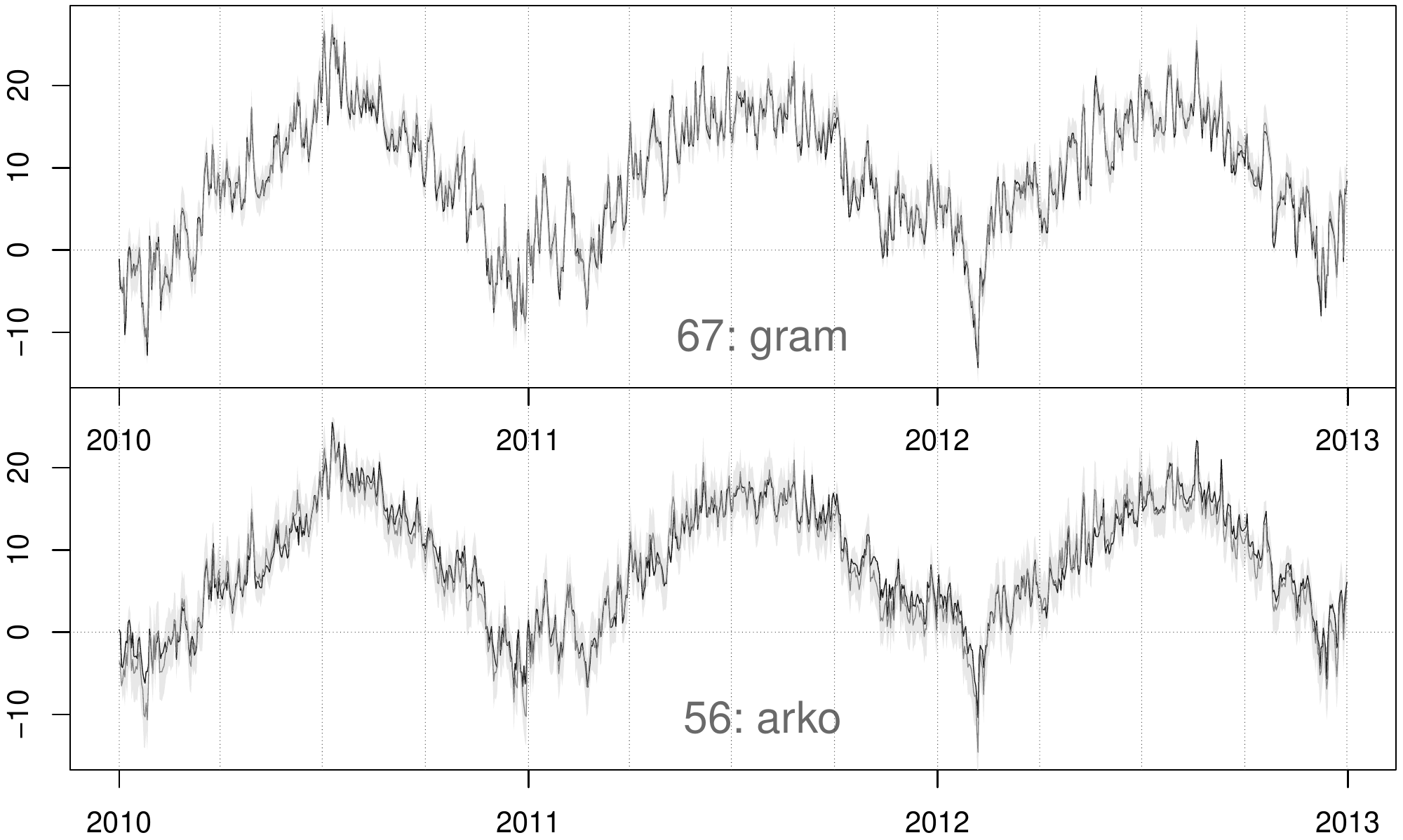}
	\caption[Prediction of the mean temperatures for the observation stations \emph{Grambek} ($67$) and \emph{Arkona} ($56$) based on the spatial Gaussian model.]{Prediction of the mean temperatures for the observation stations \emph{Grambek} ($67$) and \emph{Arkona} ($56$) for the period 01/01/2010-12/31/2012 based on the spatial Gaussian model. black line: observed values. dark gray line: prediction. light gray area: $95\%$ prediction intervals.}
	\label{fig:STpredMSE}
\end{sidewaysfigure}

\begin{sidewaysfigure}
	\centering
		\includegraphics[width=1.0\textwidth]{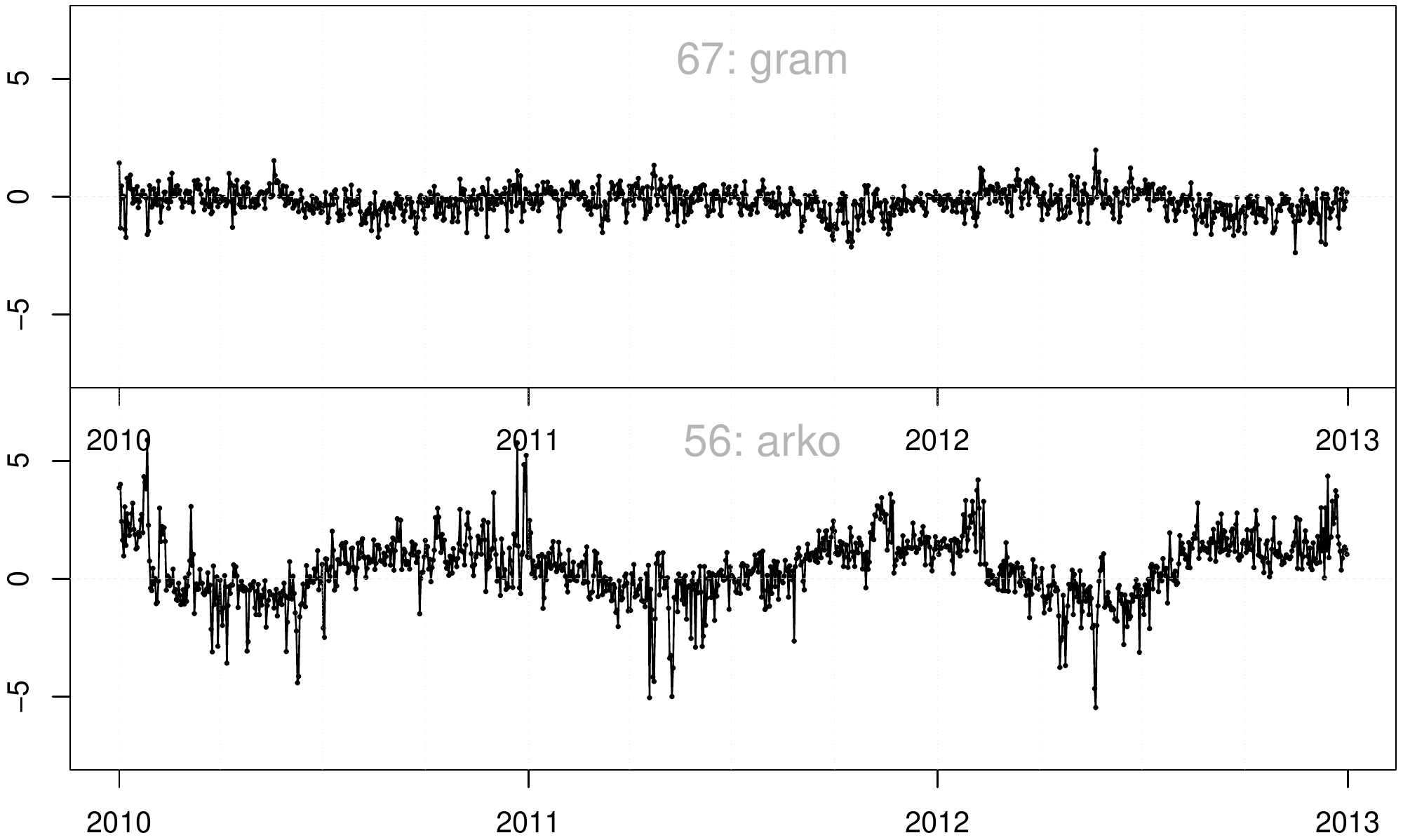}
	\caption[Prediction errors of the predictions for the observation stations \emph{Grambek} ($67$) and \emph{Arkona} ($56$) based on the spatial Gaussian model.]{Prediction errors of the predictions for the observation stations \emph{Grambek} ($67$) and \emph{Arkona} ($56$) for the period 01/01/2010-12/31/2012 based on the spatial Gaussian model.}
	\label{fig:STpredresMSE}
\end{sidewaysfigure}

\end{document}